\documentclass[
twocolumn,
superscriptaddress,
preprintnumbers,
prb,
]{revtex4-2}
\usepackage{enumitem}
\usepackage{amsmath,amssymb}
\usepackage{graphicx}% Include figure files
\usepackage{dcolumn}% Align table columns on decimal point
\usepackage{bm}% bold math
\usepackage{float}
\usepackage[dvipsnames]{xcolor}
\usepackage{hyperref}
\hypersetup{colorlinks=true,allcolors=MidnightBlue}
\usepackage{braket}  
\usepackage{xcolor}

\begin{document}

\title{Flat-plane based double-counting free and parameter free  
 many-body DFT+U}

\author{Andrew C. Burgess}
\affiliation{%
School of Physics, Trinity College Dublin, The University of Dublin, Ireland}

 \author{David D. O'Regan}%
 \email{david.o.regan@tcd.ie}
\affiliation{%
School of Physics, Trinity College Dublin, The University of Dublin, Ireland}

\date{\today}% It is always \today, today,

\begin{abstract}
Burgess et al. have recently introduced the BLOR corrective
exchange-correlation functional 
that is, by construction, the 
unique simplified rotationally-invariant  DFT$+U$ functional 
that enforces the flat-plane condition separately on each 
effective orbital of a localized subspace.
Detached  from the Hubbard model, functionals of this type are
both double-counting correction free and, 
when optimized in situ using appropriate 
error quantifiers, effectively parameter free.
They are as computationally undemanding as conventional DFT+$U$ functionals.
In this work, the extension  
of the BLOR functional to address many-body errors (mBLOR)
is derived. 
The mBLOR functional is built to enforce the flat-plane condition 
on the entire subspace, rather than  on each orbital individually. 
It depends solely on the total subspace occupancy and spin magnetization, bringing   
consistency with how Hubbard $U$ and Hund $J$ 
values are typically calculated, and very low complexity.
In this way  inter-orbital  errors are corrected on the 
same footing as the single-particle ones.
Focusing on exact test cases with strong inter-orbital interactions, 
the BLOR and mBLOR functionals were benchmarked  
against contemporary DFT+$U$ functionals using  
the total energy extensivity condition on 
stretched homo-nuclear p-block dimers that represent various 
 self-interaction and static-correlation error regimes, 
namely singlet N$_2$ \& F$_2$, non-spin polarized O$_2$, and doublet Ne$_2^+$.
The  BLOR functional outperformed all other DFT$+U$ functionals tested,
which often act to increase total-energy errors, yet it 
still yielded large  errors in some systems.
mBLOR instead yielded low 
energy errors across all four strongly-correlated dimers,
while being 
constructed using only semi-local approximation ingredients.
As mBLOR would not otherwise 
introduce a band-gap correction in the  manner 
that is a desirable feature of DFT+$U$, we developed  
a  cost-free technique to reintroduce 
it automatically  by moving the functional's unusual explicit derivative 
discontinuity into the potential.
With this in place, mBLOR is the only known 
DFT$+U$ functional that opens the bandgap of stretched 
neutral homo-nuclear dimers without the aid of
unphysical spin-symmetry breaking.

\end{abstract}

\maketitle

%\tableofcontents
Density Functional Theory (DFT)~\cite{hohenbergInhomogeneousElectronGas1964a,kohnSelfConsistentEquationsIncluding1965,m.tealeDFTExchangeSharing2022,perdewFourteenEasyLessons2010} is a central element of modern-day 
condensed matter physics and materials chemistry.
It is a leading method for the prediction of the electronic, magnetic and crystallographic structures of solid state materials thanks largely to its favorable balance between predictive accuracy and computational efficiency. This  balance strongly depends on the choice of exchange-correlation (XC) approximation. 

However, current XC approximations such as the Local Spin Density Approximations, (LSDAs)~\cite{voskoAccurateSpindependentElectron1980,barthLocalExchangecorrelationPotential1972,perdewAccurateSimpleAnalytic1992,perdewSelfinteractionCorrectionDensityfunctional1981,entwistleLocalDensityApproximations2016}, Generalized Gradient Approximations (GGAs)~\cite{perdewGeneralizedGradientApproximation1996,leeDevelopmentColleSalvettiCorrelationenergy1988,perdewRestoringDensityGradientExpansion2008,beckeDensityfunctionalExchangeenergyApproximation1988,vermaHLE16LocalKohn2017}, meta-GGAs~\cite{sunStronglyConstrainedAppropriately2015,m.delcampoNewMetaGGAExchange2012,taoClimbingDensityFunctional2003,zhaoNewLocalDensity2006,wangRevisedM06LFunctional2017} and hybrid 
functionals~\cite{beckeDensityFunctionalThermochemistry1993,perdewRationaleMixingExact1996} are all known to fail in cases where the electronic system contains one or more isolated, localized subsystems.
To mention a prototypical example, 
this failure was investigated by Perdew for the quintessential Mott insulator~\cite{perdewDensityFunctionalTheory1985}, the hydrogen lattice at the infinite atomic separation limit, which has 
an ionization energy of 13.6 eV and a a bandgap of 12.8 eV. Perdew showed that the LSDA  {underestimates} the ionization energy by approximately 46\% and, moreover, that it 
 {spuriously predicts the hydrogen lattice to be} metallic.

Such failures of the LDA, GGAs, meta-GGAs and hybrid functionals are also well documented for the dissociation of molecular dimers~\cite{ruzsinszkySpuriousFractionalCharge2006,dutoiSelfinteractionErrorLocal2006,nafzigerFragmentbasedTreatmentDelocalization2015} and in the prediction of 3d transition metal oxides, where in both cases valence electrons remain partially or completely localized on atomic sites.

\section{Double-counting free treatment of 
Localized Electronic States}
In the case of electronic systems with isolated localized subsystems, the energy of each isolated localized state should  obey the tilted plane condition~\cite{burgessTiltedPlaneStructureEnergy2024} in the limit where the interaction between the localized state and the bath (the rest of the electronic system) is negligible, or more generally, when the subspace-bath interaction energy varies linearly with spin resolved subspace occupancy. Standard XC functionals however, will exhibit spurious curvature in the energy of each isolated, localized state. This will result in erroneous total energies of the global system whenever the isolated subspaces have non-integer values of subspace electron count $N$ and subspace magnetization $M$. These errors in the total energy will occur even at integer values of global electron count $N_{\rm tot}$ and magnetization $M_{\rm tot}$~\cite{mori-sanchezDerivativeDiscontinuityExchange2014}.
We refer to these energy errors as localized many electron self interaction errors and localized static correlation errors~\cite{burgessMathrmDFTTexttypeFunctional2023,zhaoGlobalLocalCurvature2016} to distinguish them 
from their global analogues~\footnote{The term local could
also be used here, but in the DFT context it seems preferable
to reserve `local' to describe variables that depend 
explicitly upon a single spatial argument.}.

Noting the qualitative failures of standard XC approximations when  simulating Mott insulators, Anisimov et al.~\cite{anisimovBandTheoryMott1991} proposed supplementing such XC approximations with an additional electronic interaction term inspired by the Hubbard model~\cite{hubbardElectronCorrelationsNarrow1963},
\begin{equation}
\label{eqn:hubbard_interaction}
E_{\rm int}=\frac{U}{2}\sum_{m m' \sigma}n_{m\sigma}n_{m'\bar{\sigma}}+\frac{U-J}{2}\sum_{mm'\sigma \atop m\neq m'}n_{m\sigma}n_{m'\sigma},
\end{equation}
which explicitly depends on the spin resolved occupancy $n_{m \sigma}$ of each orbital $m$ in the localized subspace. $U$ and $J$ refer to the Hubbard $U$ and Hund's $J$ interaction parameters which need to be chosen or preferably evaluated~\cite{linscottRoleSpinCalculation2018,cococcioniLinearResponseApproach2005,timrovHubbardParametersDensityfunctional2018,lambertUseMathrmDFT2023,pickettReformulationMathrmLDA1998,springerFrequencydependentScreenedInteraction1998,moseyInitioEvaluationCoulomb2007,moseyRotationallyInvariantInitio2008}, in advance.  The inter-electron interactions of Eq.~\ref{eqn:hubbard_interaction} however, are already accounted for to a less favourable extent by the original XC functional that the Hubbard-like interaction term is designed to supplement. This necessitates the use of a double counting correction scheme. The inter-electron interaction and double counting scheme together form a DFT$+U$-type functional~\cite{anisimovBandTheoryMott1991,anisimovDensityfunctionalTheoryNiO1993,czyzykLocaldensityFunctionalOnsite1994,pickettReformulationMathrmLDA1998,liechtensteinDensityfunctionalTheoryStrong1995,dudarevElectronenergylossSpectraStructural1998,seoSelfinteractionCorrectionMathrmLDA2007,himmetogluFirstprinciplesStudyElectronic2011,bajajCommunicationRecoveringFlatplane2017,bajajNonempiricalLowcostRecovery2019,shishkinDFTDudarevFormulation2019,dudarevParametrizationMathrmLSDANoncollinear2019,burgessMathrmDFTTexttypeFunctional2023,campoExtendedDFTMethod2010,petukhovCorrelatedMetalsMathrm2003,solovyevCorrectedAtomicLimit1994,floresClosedLocalorbitalUnified2022,parkDensityFunctionalSpindensity2015}. Over the past three decades numerous double counting correction schemes have been developed including the spin polarized and non-spin polarized analogues of the around mean field~\cite{anisimovBandTheoryMott1991,czyzykLocaldensityFunctionalOnsite1994} and fully localized limit~\cite{anisimovDensityfunctionalTheoryNiO1993,czyzykLocaldensityFunctionalOnsite1994} double counting schemes. In practical applications, the predicted physical and chemical properties of a material can strongly depend on the choice of double counting scheme~\cite{ylvisakerAnisotropyMagnetismText2009,ryeeEffectDoubleCounting2018}. Despite this strong dependency, DFT$+U$-type functionals enjoy widespread application, most notably in high throughput material screening~\cite{mooreHighthroughputDeterminationHubbard2024a,bennettSystematicDeterminationHubbard2019,hortonHighthroughputPredictionGroundstate2019,hegdeQuantifyingUncertaintyHighthroughput2023}, which can require thousands of DFT$+U$-type calculations to be executed.

Developing a DFT$+U$-type functional that yields reliable total energies and bandgaps is thus of crucial importance.
Improved understanding in recent years of exact conditions that 
apply to the atomic limit has given the opportunity to design 
DFT$+U$ that encode those, without adding complexity, rather
than invoking approximate double-counting corrections~\cite{burgessMathrmDFTTexttypeFunctional2023}.
Meanwhile, incorporation from the field of quantum chemistry of 
the use of test systems for which exact total energies, 
or at least exact total-energy differences, are available, 
enables the stringent determination of the viable 
class of functional forms of a given complexity level~\cite{burgessMathrmDFTTexttypeFunctional2023,bajajCommunicationRecoveringFlatplane2017,bajajNonempiricalLowcostRecovery2019}.
Departing entirely from the Hubbard model and hence circumventing
the introduction of a double-counting correction, in this work we develop a DFT$+U$-type functional whose 
form is entirely based upon exact conditions in DFT.
Specifically, we focus on the regime in which 
the DFT$+U$ subspaces are relatively weakly interacting
with the remainder of the system, so that they harbor the 
exact flat-plane (or more generally tilted-plane) condition, 
and where their deviations from that condition,
many-body self-interaction error (SIE) and many-body static-correlation
error (SCE), are treated in a subspace-averaged fashion in the 
usual pragmatic, cost-effective manner of DFT+$U$.

\section{Exact Conditions in DFT}
The poor performance of most XC functionals in the prediction of electronic systems with partially or completely localized states, can be largely attributed to their breaking of certain physical conditions~\cite{kaplan2023predictive} that the exact XC functional is known to obey, the best known among which is the piecewise linearity condition with respect to electron count~\cite{perdewDensityFunctionalTheoryFractional1982,yangDegenerateGroundStates2000a,ayersDependenceContinuityEnergy2008}. The total ground-state energy of a system with non-integer electron count $N_{\rm tot}$, must be a linear interpolation of the ground-state energies of the same system with integer values of electron count $N_0$ and $N_0+1$,
\begin{align}
&E_v[N_{\rm tot}] =\omega E_v[N_0]+(1-\omega)E_v[N_0+1],   \\
&{\rm where } \quad N_{\rm tot} = \omega N_0+(1-\omega)(N_0+1) , \quad 0 \leq \omega \leq 1. \nonumber
\end{align}
This exact condition was originally explicated by Perdew et al.~\cite{perdewDensityFunctionalTheoryFractional1982} assuming the convexity condition with respect to electron count is satisfied,
\begin{equation}
\label{eqn:convexity}
2E_v[N_{\rm tot}] \leq E_v[N_{\rm tot}-1]+E_v[N_{\rm tot}+1].
\end{equation}
Recently the convexity condition has been proven 
to be true for all electronic systems within DFT~\cite{burgessConvexityConditionDensityfunctional2023}, 
whenever the  density functional is (1) exact for all $v-$representable densities, (2) size-consistent, 
and (3) translationally invariant. 

\begin{figure}
\centering
\includegraphics[scale=0.6]{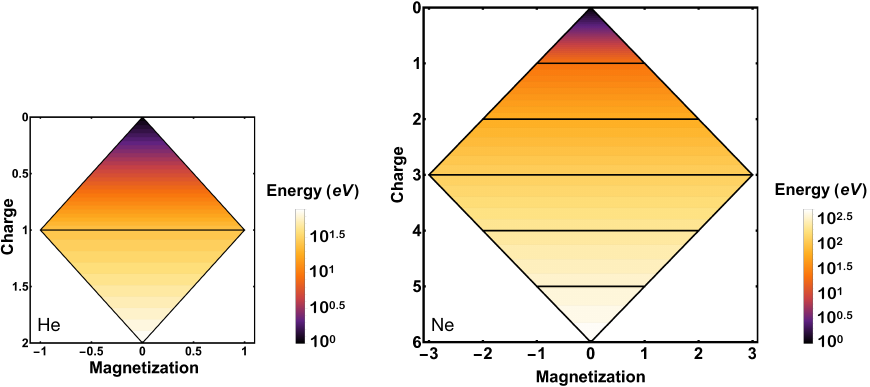}
\caption{The projection of the $E_v[N_{\rm tot},M_{\rm tot}]$ curve of the helium atom (left) and neon atom (right) onto the $N_{\rm tot}-M_{\rm tot}$ plane. For simplicity, highly charged and highly magnetized states are omitted. The total energy varies linearly across each plane, which are outlined in black. The total energy values are based on available experimental NIST reference data \cite{NIST_ASD} and are given relative to the lowest energy state of the respective neutral atom which is set to 1 eV. Energy contour lines are plotted at intervals of $10^{0.05x}$~eV. In the BLOR functional~\cite{burgessMathrmDFTTexttypeFunctional2023},
the spin-dependent occupancy of each DFT+$U$ subspace 
spatial orbital can be  {thought} of as being separately 
mapped, conceptually,
to its own possibly-tilted helium atom (left) exact energy model.
In the introduced mBLOR functional, and taking the example
of p-orbital based subspaces, the spin-dependent occupancies
of all three subspace spatial orbitals are together mapped
to a combined possibly-tilted neon 2p subshell 
(right) exact energy model.}
\label{fig:neon_energy_curve}
\end{figure}

An analogous exact condition of equal importance is the piecewise linearity condition with respect to magnetization~\cite{burgessTiltedPlaneStructureEnergy2024,galEnergySurfaceChemical2010,cohenFractionalSpinsStatic2008}. The total ground-state energy of a system with non-integer magnetization $M_{\rm tot}$ but integer electron count $N_0$, must be a linear interpolation of the ground-state energies of the same system with integer values of magnetization $M_i$ and $M_j$,
\begin{align}
\label{eqn:magnetization_set_up}
&E_v[N_0,M_{\rm tot}]=\omega E_v[N_0, M_i]+(1-\omega)E_v[N_0, M_j],  \\ 
&M_{\rm tot}=\omega M_i+(1-\omega)M_j, {\rm \hspace{0.3cm}} M_i, M_j \in \mathbb{Z} {\rm \hspace{0.3cm}} \& {\rm \hspace{0.3cm}} 0 \leq \omega \leq 1. \nonumber
\end{align}
These two exact conditions have been combined and extended to give the flat plane condition~\cite{chanFreshLookEnsembles1999,yangDegenerateGroundStates2000a,mori-sanchezDiscontinuousNatureExchangeCorrelation2009a,perdewExactExchangecorrelationPotentials2009,devriendtQuantifyingDelocalizationStatic2021,devriendtUncoveringPhaseTransitions2022,g.janeskoReplacingHybridDensity2021,yangCommunicationTwoTypes2016,cuevas-saavedraSymmetricNonlocalWeighted2012,galEnergySurfaceChemical2010,goshenEnsembleGroundState2024,richerSpinPolarizedConceptualDensity2024,richerSpinPolarizedConceptualDensity2024}, or more generally the tilted plane condition~\cite{burgessTiltedPlaneStructureEnergy2024,malekDiscontinuitiesEnergyDerivatives2013} when the full range of possible magnetization states are considered. The tilted plane condition requires that the exact $E_v[N_{\rm tot},M_{\rm tot}]$ curve for any finite electronic system is composed of a tiled surface with vertices occurring at certain integer values of electron count and magnetization,  {with} derivative discontinuities along the edges of each tile.  Several functionals have been developed that either fully or partially satisfy the flat or tilted plane condition ~\cite{burgessMathrmDFTTexttypeFunctional2023,bajajCommunicationRecoveringFlatplane2017,bajajNonempiricalLowcostRecovery2019,janeskoMulticonfigurationalCorrelationDFT2024,suDescribingStrongCorrelation2018,proynovCorrectingChargeDelocalization2021,kongDensityFunctionalModel2016,johnsonCommunicationDensityFunctional2011,prokopiouOptimalTuningPerspective2022,cuevas-saavedraSymmetricNonlocalWeighted2012,zhouUnifiedTreatmentDerivative2018,vanaggelenExchangecorrelationEnergyPairing2013}, or help mitigate deviations from it. The exact flat plane condition for the helium
atom and neon atom is illustrated in Fig.~\ref{fig:neon_energy_curve}.

LDA, GGAs, meta-GGAs and hybrid functionals do not obey the tilted plane condition and instead exhibit a spurious curvature in the total energy with respect to $N_{\rm tot}$ and $M_{\rm tot}$. In the literature, this erroneous behaviour is typically referred to as Many Electron Self Interaction Error~\cite{ruzsinszkySpuriousFractionalCharge2006,mori-sanchezManyelectronSelfinteractionError2006} and Static Correlation Error~\cite{cohenFractionalSpinsStatic2008},  respectively.

\section{The BLOR flat-plane condition based 
DFT+U Functional}
Burgess, Linscott, and O'Regan recently developed the BLOR DFT$+U$ functional~\cite{burgessMathrmDFTTexttypeFunctional2023}, which is  defined to enforce the flat plane condition on each orbital of a localized subspace, thus dispensing with the need to invoke the Hubbard model or a double counting correction scheme.
The functional form of BLOR is uniquely specified by its
definition,  {as proven in Supplementary Material S-II 
of Ref.~\onlinecite{burgessMathrmDFTTexttypeFunctional2023}.}
In order to enforce the localized tilted plane condition, the spin resolved Hubbard $U^{\sigma}$  parameter must be defined as the spurious curvature (in the physics
sense meaning second derivative) of an appropriatetly defined 
energy with respect to spin resolved 
subspace occupancy $n^{\sigma}$ (at fixed $n^{\bar{\sigma}}$,  {where
$\bar{\sigma}$ is the spin opposite 
to $\sigma$.}). Similarly the magnitude of the Hund's $J$ parameter must be defined as the spurious curvature of an appropriate  {energy} with respect to 
subspace spin-magnetization (at fixed subspace occupancy). 
We note that localized spurious energy curvatures with respect to 
charge are  closely related, albeit somewhat different technically, to
the violation of the generalized DFT Koopmans' theorem endemic to  
practical approximate functionals.
The currently most practical comprehensive remedy for the latter, in 
in-situ corrective form, are  
the Koopmans compliant functionals detailed, e.g., in Refs.~\onlinecite{daboKoopmansConditionDensityfunctional2010,borghiKoopmanscompliantFunctionalsTheir2014,colonnaKoopmansSpectralFunctionals2022}.

The BLOR functional was found to yield highly accurate total energies for s-valence molecular systems at large separation lengths, namely H$_2$, He$_2^+$, Li$_2$, Be$_2^+$ and the triplet H$_5^+$ ring, with relative energy errors below 0.6\%. These same molecular systems suffer from relative energy errors as high as 8.0\% at the raw DFT  {level}, using the Perdew-Burke-Ernzerhof (PBE) approximation. Use of current standard DFT$+U$-type functionals such as Dudarev's 1998 Hubbard functional~\cite{dudarevElectronenergylossSpectraStructural1998} or Liechtenstein's 1995 Hubbard functional~\cite{liechtensteinDensityfunctionalTheoryStrong1995} (the two corrective functionals are equivalent in the case of s-valence species) often significantly worsened the raw 
approximate DFT (PBE functional) total energy, with relative energy errors as high as 20.5\%. In the case of multi-orbital subspaces, the BLOR functional was designed to enforce the localized tilted plane condition separately on each orbital of the subspace,  {and is
defined by}
\begin{widetext}
\begin{align}
\label{eqn:BLOR_trace_version}
E_{\rm BLOR}= \left\{
\begin{array}{*6{>{\displaystyle}c}}
%{\rm \hspace{8mm}}
\frac{U^{\upharpoonright}+U^{\downharpoonright}}{4}{\rm Tr}[\hat{N}-\hat{N}^2]
%{\rm \hspace{10.5mm}}
&+&
%{\rm \hspace{10.5mm}}
\frac{J}{2}{\rm Tr}[\hat{M}^2-\hat{N}^2]
%{\rm \hspace{6.75mm}}
&+&
%{\rm \hspace{6.75mm}}
\frac{U^{\upharpoonright}-U^{\downharpoonright}}{4}{\rm Tr}[\hat{M}-\hat{N}\hat{M}],
%{\rm \hspace{5mm}}
&  \
%{\rm \hspace{5mm}}
{\rm Tr}[\hat{N}] \leq {\rm Tr}[\hat{P}].
 \\
{\frac{U^{\upharpoonright}+U^{\downharpoonright}}{4}{\rm Tr}[(\hat{N}-\hat{P})-(\hat{N}-\hat{P})^2]}
%{\rm \hspace{2mm}}
&+&
%{\rm \hspace{2mm}}
{\frac{J}{2}{\rm Tr}[\hat{M}^2-(\hat{N}-2\hat{P})^2]}
%{\rm \hspace{2mm}}
&+&
%{\rm \hspace{2mm}}
{\frac{U^{\upharpoonright}-U^{\downharpoonright}}{4}{\rm Tr}[\hat{M}-\hat{N}\hat{M}]},
%{\rm \hspace{5mm}}
&  \
%{\rm \hspace{5mm}}
{\rm Tr}[\hat{N}] > {\rm Tr}[\hat{P}],
\end{array}
\right.
\end{align}
\end{widetext}
where the subspace occupancy and magnetization operators can be
expressed in terms of the spin-resolved subspace occupancy
operators $\hat{N} = \hat{n}^{\upharpoonright} + \hat{n}^{\downharpoonright}$ and $\hat{M} = \hat{n}^{\upharpoonright} - \hat{n}^{\downharpoonright}$. The spin-resolved subspace occupancy operator  {is}  $\hat{n}^{\sigma} = \hat{P}\hat{\rho}^{\sigma}\hat{P}$, where $\hat{\rho}^{\sigma}$ is the spin-$\sigma$ Kohn-Sham density operator and $\hat{P}$ is the subspace projection operator. 

By enforcing the tilted plane condition separately on each orbital of the subspace, the BLOR functional acts to enforce a helium like flat plane condition, as depicted in Fig.~\ref{fig:neon_energy_curve}, on each of the Tr$[\hat{P}]$ orbitals of the subspace. Following
standard practice in DFT+U, this orbital resolved functional is then applied using subspace-averaged $U$ and $J$ parameters. This amounts to an 
approximation or inconsistency, 
but perhaps more importantly to the neglect of 
inter-orbital corrections within the subspace, 
and hence the restriction to  
single-orbital rather than many-body 
SIE and SCE. 

The alternative and complimentary approach, pursued in this study, is to apply the subspace resolved $U$ and $J$ parameters using a subspace resolved DFT$+U$-type functional, which then depends only the total subspace occupancy and magnetization. In this case, the DFT$+U$ functional effectively enforces one neon-like flat plane condition (for the example of a p-orbital subspace;
for d-orbitals we would refer to zinc) as opposed to enforcing a set of Tr$[\hat{P}]$ separate helium like flat plane conditions.
Before explicating this alternative DFT$+U$-type functional in full, we will first elaborate on how these subspace averaged $U$ and $J$ parameters are evaluated. 

\section{Hubbard and Hund Parameters}
Measuring the spurious curvature in the energy with respect to occupancy, while fixing the spin-magnetization, would in 
principle require cumbersome and often ill-conditioned
self-consistent constrained DFT calculations. 
Instead, one can directly evaluate the corrective parameters 
$U$ and $J$ 
based on first-order partial derivatives of the spin-resolved, subspace averaged Hartree and exchange-correlation (Hxc) potential,
\begin{equation}
v_{Hxc}^{\sigma}={\rm Tr}\left[\hat{P}{\rm \enspace}\frac{\delta E_{\rm Hxc}^{\rm total}}{\delta \hat{\rho}^{\sigma}}\right]\bigg/{\rm Tr}[\hat{P}],
\end{equation}
where $E_{\rm Hxc}^{\rm total}$ is the total Hxc energy of the system and $\hat{P}$ is the subspace projection operator. Using $v_{Hxc}^{\sigma}$, we can define the spin-resolved subspace Hxc interaction
\begin{equation}
f^{\sigma\sigma'}=\left.\frac{\partial v_{\rm Hxc}^{\sigma}}{\partial n^{\sigma' }}\right|_{n^{\bar{\sigma}'}},
\end{equation}
where $n^{\sigma' }$ is the spin $\sigma'$ subspace occupancy. This is known as the minimum-tracking linear response method~\cite{linscottRoleSpinCalculation2018,moynihanSelfconsistentGroundstateFormulation2017,orhanFirstprinciplesHubbardHund2020,bermanReconcilingTheoreticalExperimental2023}. 
Within this formalism, 
the spin resolved Hubbard parameters for BLOR 
(or mBLOR) 
can be set as the diagonal elements of the Hxc interaction, 
namely as 
\begin{equation}
\label{eqn:Usigma}
U^{\sigma}=f^{\sigma \sigma }.
\end{equation}
The Hund's $J$ parameter is defined as the spurious curvature in the interacting part of the energy with respect to magnetization. 
Within the minimum-tracking linear response 
formalism it is normally defined as 
\begin{equation}
J=-\frac{1}{2}\frac{d v_{\rm Hxc}^{\upharpoonright}-d v_{\rm Hxc}^{\downharpoonright}}{d(n^{\upharpoonright}-n^{\downharpoonright})}.
\end{equation}
However, this derivative with respect to magnetization should,  {for
functionals such as BLOR 
and mBLOR,} be evaluated at fixed total occupancy. This variation
can be calculated very conveniently 
using the simple $2\times 2$ method~\cite{linscottRoleSpinCalculation2018}, where it is given by 
\begin{equation}
\label{eqn:Jsimple}
J=-\frac{1}{4}(f^{\upharpoonright \upharpoonright}-f^{\upharpoonright \downharpoonright}-f^{\downharpoonright \upharpoonright}+f^{\upharpoonright \downharpoonright}). 
\end{equation}
Similarly, within the minimum-tracking linear response formalism the spin-independent Hubbard $U$ parameter is usually defined as
\begin{equation}
U=\frac{1}{2}\frac{d v_{\rm Hxc}^{\upharpoonright}+d v_{\rm Hxc}^{\downharpoonright}}{d(n^{\upharpoonright}+n^{\downharpoonright})}
\end{equation}
and, again, there exists an analogous simple $2\times 2$ 
variant (particularly, but not necessarily only 
for use with flat-plane condition based functionals),
\begin{equation}
\label{eqn:Usimple}
U=\frac{1}{4}(f^{\upharpoonright \upharpoonright}+f^{\upharpoonright \downharpoonright}+f^{\downharpoonright \upharpoonright}+f^{\upharpoonright \downharpoonright}).
\end{equation}
Thus, the Hubbard $U$ (and Hund's $J$) parameters are typically evaluated as subspace averaged localized 
many-body SIE (and SCE) strengths, or subspace averaged
interaction strengths of a specfic type, 
depending on the reader's perspective. Despite this subspace averaging, the $U$ and $J$ parameters are typically employed in orbitally resolved DFT+U-type functionals, such as the original BLOR functional of Eq.~\ref{eqn:BLOR_trace_version} or Dudarev et al.'s functional~\cite{dudarevElectronenergylossSpectraStructural1998},
\begin{equation}
E_u=\frac{U-J}{2}\sum_{\sigma m m'} n^{\sigma}_{mm'}\delta_{mm'}-n^{\sigma}_{mm'}n^{\sigma}_{m'm}.
\end{equation}
These subspace averaged $U$ and $J$ parameters can in principle be decomposed into orbitally resolved contributions
\begin{equation}
f^{\sigma \sigma'}_{mm'}=\bigg(\frac{\partial v_{{\rm Hxc}}^{m \sigma}}{\partial n_{m' \sigma'}}\bigg),
\end{equation}
although we emphasize  that this decomposition is not performed in this study. In an atomic subspace of orbital angular momentum quantum number $l$, for every one on-diagonal intra-orbital term $f^{\sigma \sigma'}_{mm}$ contributing to the subspace averaged $U$ and $J$ parameters there will be $2l$ off-diagonal inter-orbital terms $f^{\sigma \sigma'}_{mm'}$, where $m\neq m'$. Due to the contributions from these inter-orbital terms $f^{\sigma \sigma'}_{mm'}$, it is inconsistent  
to use such subspace averaged $U$ and $J$ parameters in an orbitally resolved DFT+ U-type functional. 
This approximation will tend to ascribe all the measured many-body localized SIE and SCE to the inter-orbital (single-particle)  terms,
likely over-correcting them, while neglecting the numerous
inter-orbital corrective terms. How great an issue this is in 
practice is likely dependent on the system under study and
the choice of subspace projection. It seems likely to be more
problematic, for the accuracy of the total energy, when there
are more than one significantly partially occupied subspace
spin-orbitals. 
This inconsistency can, of course,  be resolved by following the long-known route of evaluating orbitally resolved corrective parameters~\cite{pickettReformulationMathrmLDA1998,mackeOrbitalResolvedDFTMolecules2024,meiExactSecondOrderCorrections2021,liLocalizedOrbitalScaling2018} or alternatively, one could develop a non-orbitally resolved DFT$+U$-type functional designed to mitigate errors or account for on-site interactions associated with the subspace as a whole as opposed to each orbital separately. It is the later of these two options which we choose to pursue in this study. Unlike the orbitally resolved parameter approach, this one does not present challenges in maintaining a functional form that is invariant under unitary transformations of the orbitals. In principle, this 
latter approach could be applied to subspaces that are not orbital based at all, 
 {but for example
defined by a weight function in real space, 
which might prove most helpful} in orbital-free DFT.

\section{Flat-plane based functional 
that includes 
inter-orbital corrections for many-body errors: mBLOR in 
the spin-symmetric case}
We are now ready to derive the generalization of the BLOR functional to address  many-body SIE and SCE, termed mBLOR for brevity. For simplicity, we will first consider the simpler case of a spin symmetric system whose total subspace energy satisfies 
\begin{equation}
\label{eqn:spin_degeneracy}
E[N,M]=E[N,-M].
\end{equation}
Here, $N$ and $M$ are the total subspace electron count and subspace spin-magnetization 
(note here that magnetization is measured in units of electrons, 
following convenient convention in DFT, not units of $\hbar$),
respectively,
\begin{equation}
N=\sum_{\sigma m }n^{\sigma}_{mm}\quad \& \quad M=\sum_{ m }(n^{\upharpoonright}_{mm}-n^{\downharpoonright}_{mm}).
\end{equation}
This spin-symmetric special case simplifies matters considerably because  the spurious curvature in the total subspace energy of 
an approximate XC functional with respect to spin up subspace occupancy $n^{\upharpoonright}$ is then  equal to the spurious curvature in the total subspace energy with respect to spin down subspace occupancy $n^{\downharpoonright}$. Furthermore, if $[N,M]$ is a vertex in the energy landscape, then $[N,-M]$ is a degenerate vertex in the energy landscape. 

Focusing on the lowest energy magnetization states, 
and assuming no structure to the weak interaction with 
the bath for the subspace,
the $E[N,M]$ surface is as shown in Fig.~\ref{fig:neon_energy_curve}
(for the cases of single s or p-valence subspaces, the
d and f cases being straightforward generalizations)
assuming two criteria are satisfied. 
\begin{enumerate}[label=(\alph*)]
\item 
The total energy of the subspace is strictly convex with respect to electron count, so that 
\begin{equation}
2E[N]< E[N-1]+E[N+1].
\end{equation}

    \item 
Hund's First Rule is satisfied, i.e., each orbital of the subspace is occupied singly first with electrons of parallel spin, before any double occupation occurs. 
\end{enumerate}
As shown in Fig.~\ref{fig:neon_energy_curve}, the $E[N,M]$ surface in this case is composed of a series of flat planes, which meet with derivative discontinuities at integer values of electron count. The $E[N,M]$ surface in Fig.~\ref{fig:neon_energy_curve}, displays a large number of vertices. In general we expect the approximate XC functional, that BLOR is designed to supplement, to yield accurate total energies for atomic systems whenever $N$ and $M$ are located at vertices in the energy landscape. It is important to note that an analogous $E[N,M]$ surface to Fig.~\ref{fig:neon_energy_curve} will also occur for d and f valence-atoms if criteria (a)-(b) are satisfied. 

\begin{figure*}
  \includegraphics[width=\textwidth]{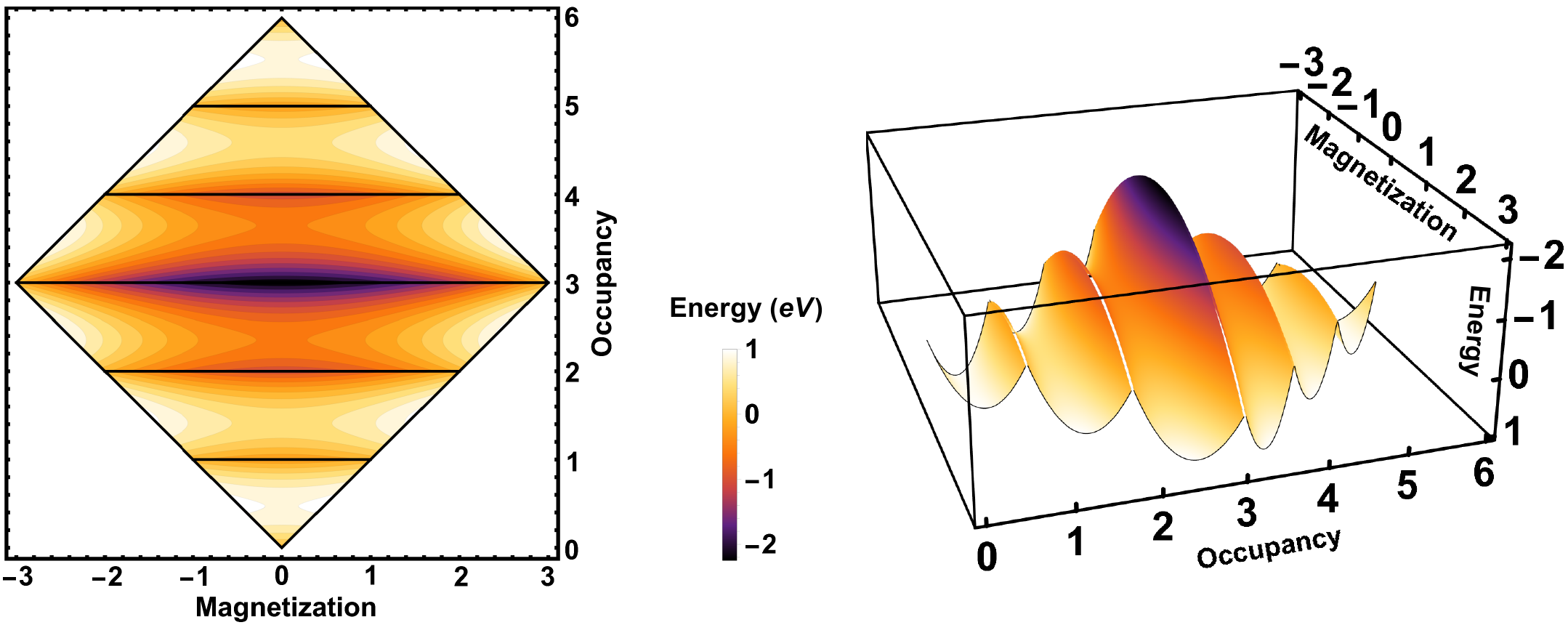}
\caption{Plot of the total corrective energy of the spin-symmetric mBLOR functional of Eq.~\ref{eqn:mBLORsym} for a p-orbital subspace with Hubbard corrective parameters  {$U=$ 8.5} eV and $J=$ 0.5 eV. In the left-hand image energy contour lines are plotted at intervals of $0.2$ eV. In the right hand image a side-on view of the same mBLOR total corrective energy surface is presented.}
\label{fig:subspace_fracturing}
\end{figure*}

In order to enforce the tilted plane condition once  {on} a subspace that satisfies criteria (a)-(b), the corrective functional must satisfy four key conditions. It must: 
\begin{enumerate}
    \item 
Be a continuous function of the subspace electron
count $N$ and magnetization $M$.
    \item 
Yield no correction at the vertices in the $E[N,M]$ landscape. This
is desirable because approximate functionals are expected to
yield accurate total energies in this case.
    \item 
Have a constant curvature of $-U$ with respect to $N$. This is desirable because approximate functionals are expected to have a spurious curvature of $U$ with respect to $N$ due to their deviation from the tilted plane condition, and this should be subtracted off.
    \item 
Have a constant curvature of $J$ with respect to $M$. This
is desirable because approximate functionals are expected to
have a spurious curvature of $-J$ with respect to $M$
(the minus is due to long-standing convention for defining $J$), 
and this should be subtracted off.
\end{enumerate}
The spin-symmetric version of the mBLOR functional is the 
mathematically unique functional that satisfies these four conditions for a subspace that meets criteria (a)-(b),
and it may be written separately for each subspace as 
\begin{widetext}
\begin{align}
\label{eqn:mBLORsym}
E_{\rm mBLOR}= \left\{
\begin{array}{*6{>{\displaystyle}c}}
%{\rm \hspace{8mm}}
\frac{ {U-J}}{2}\left[{(N-N_0)}-{(N-N_0)}^2\right]
%{\rm \hspace{10.5mm}}
&+&
%{\rm \hspace{10.5mm}}
\frac{J}{2}\left[{M}^2-{N}^2\right],
%{\rm \hspace{6.75mm}}
&  \
%{\rm \hspace{5mm}}
N \leq {\rm Tr}[\hat{P}],
 \\ \\
\frac{ {U-J}}{2}\left[{(N-N_0)}-{(N-N_0)}^2\right]
%{\rm \hspace{2mm}}
&+&
%{\rm \hspace{2mm}}
\frac{J}{2}\left[{M}^2-{(N-2{\rm Tr}[\hat{P}])}^2\right],
%{\rm \hspace{2mm}}
&  \
%{\rm \hspace{5mm}}
N > {\rm Tr}[\hat{P}].
\end{array}
\right.
\end{align}
\end{widetext}
In Eq.~\ref{eqn:mBLORsym} $N_0$ is defined as $\lfloor N \rfloor$, where $\lfloor \cdot \rfloor$ is the floor function (the integer part). The first term on the right hand side of Eq.~\ref{eqn:mBLORsym} is the many-electron self interaction error term. It mitigates the localized analogue of MSIE by removing a quadratic term in $N$ of curvature  {$U-J$
and replacing it with a linear term.
We have used in Eq.~\ref{eqn:mBLORsym} the fact  that
$( U^\upharpoonright + U^\downharpoonright)/2=U-J$ for spin-symmetric systems within the
simple $2 \times 2$ scheme.}  As expected, the MSIE term offers no correction to the total energy for systems with integer subspace occupancy. The second term on the right hand side of Eq.~\ref{eqn:mBLORsym} is the static correlation error correction term, which offers no energy correction for a maximally spin polarized subspace and reaches its maximum energy correction at $M=0$. The SCE term takes a different form depending on whether the subspace is more or less than half occupied, this is a consequence of satisfying Hund's First Rule. We refer to these two forms as the `early' and `late' versions of the mBLOR functional, which should be used when $N \leq {\rm Tr}[\hat{P}]$, and $N > {\rm Tr}[\hat{P}]$, respectively. Beyond half occupancy, the maximum magnetization, i.e., the value of $M$ for which the SCE term vanishes, decreases with increasing $N$ as the orbitals of the subspace are now being filled by a second electron of the opposite spin.  {A plot of the spin-symmetric mBLOR corrective energy surface as a function of subspace occupancy and magnetization is presented in Fig.~\ref{fig:subspace_fracturing}.}

In passing, we note that in addition to its primary application of treating localized spin-symmetric subspaces, the mBLOR functional of Eq.~\ref{eqn:mBLORsym} can also be used  {to} mitigate approximate XC functional's deviation from the {\it global} flat plane condition. In this case $N$ and $M$ of Eq.~\ref{eqn:mBLORsym} would be replaced by the total electron count and magnetization of the finite electronic system of interest.

\section{Spin-Symmetric Molecular Test Systems}
We tested the spin-symmetric mBLOR corrective functional of Eq.~\ref{eqn:mBLORsym} on three spin-symmetric homo-nuclear dimers at large inter-nuclear separation lengths, namely N$_2$, F$_2$ and (simulating its non-spin polarized triplet ground state)
O$_2$. For these stretched, neutral homo-nuclear dimers, an approximately integer number of electrons localizes on each atomic site. An ambiguity therefore arises as to which integer value to use for $N_0$ in the mBLOR functional. Following the precedent set in the study of the stretched s-block dimers~\cite{burgessMathrmDFTTexttypeFunctional2023}, we choose to set $N_0=2$, $3$ and $4$ for the stretched N$_2$, O$_2$ and F$_2$ molecules, respectively, so that the MSIE term in the mBLOR functional is equal to zero when the subspace occupancy is equal to that of the neutral atom and `plus one' cation. This choice in values of $N_0$ results in the `early' 
($N \le \rm{Tr} [\hat{P}]$) version of the mBLOR functional being applied to the stretched N$_2$ system and the `late' 
($N > \rm{Tr} [\hat{P}]$)
version of the mBLOR functional being applied to the stretched O$_2$ and F$_2$ systems. The corresponding versions of the BLOR functional is applied to the stretched N$_2$ molecule and to the stretched O$_2$ and F$_2$ molecules. In the case of the O$_2$ molecule, to avoid converging to a broken-symmetry spin-polarized ground state, the occupancy of the two frontier spin up and spin down  {GKS} orbitals were permitted, as necessary, to be degenerate with occupancy $0.5$. At large inter-nuclear separation lengths, a close to integer number of electrons will localise on each atomic site, therefore these systems will suffer from negligible localized-MSIE. Being non-spin polarized molecules, the magnetization at each of the atomic sites will be equal to zero, thus for a given value of $N$, these systems are located at the point of maximum localized SCE. Assuming a positive in-situ measured value of $J$, as indeed transpires to be the case,
the raw uncorrected PBE functional will thus overestimate the total energies of these stretched X$_2$ systems. 

At the dissociated limit the total energy of the X$_2$ molecule,
the energy extensivity condition holds, namely 
\begin{equation}
\label{eqn:energy_dimer}
E[{\rm X}_2]=2E[{\rm X}].
\end{equation}
However, due to localized MSIE and localized SCE, the total energy of the X$_2$ molecule at large inter-nuclear separation lengths will not be equal to twice the energy of the X atom when evaluated using an approximate XC functional. The difference between these two quantities is thus an intrinsic error associated with the approximate XC functional. By contrast, the raw PBE total energy of the isolated X atom will not  {significantly} suffer from localized MSIE or localized SCE as the system will be located at a vertex of the $E[N,M]$ energy surface by virtue of it being in the lowest energy magnetization state of the atom for a given integer value of electron count. Twice the PBE total energy of the X atom, can thus be used as our reference value. Ideally, application of a Hubbard-type corrective functional to the stretched X$_2$ system should mitigate this intrinsic energy error. The  {provided} bar charts present the energy error associated with each DFT+$U$-type functional,
giving what are close to ideal application conditions
(i.e., near the atomic limit). We define this energy error here as the energy difference between the stretched X$_2$ molecule evaluated with a given DFT(PBE)+$U$-type functional compared to twice the PBE total energy of the X atom. Using Eq.~\ref{eqn:energy_dimer} to define the exact total energy 
provides cancellation errors in the total energy 
due to the use of (very carefully designed) pseudo-potentials. 

\begin{figure}%[H]
\centering
\includegraphics[scale=0.45]{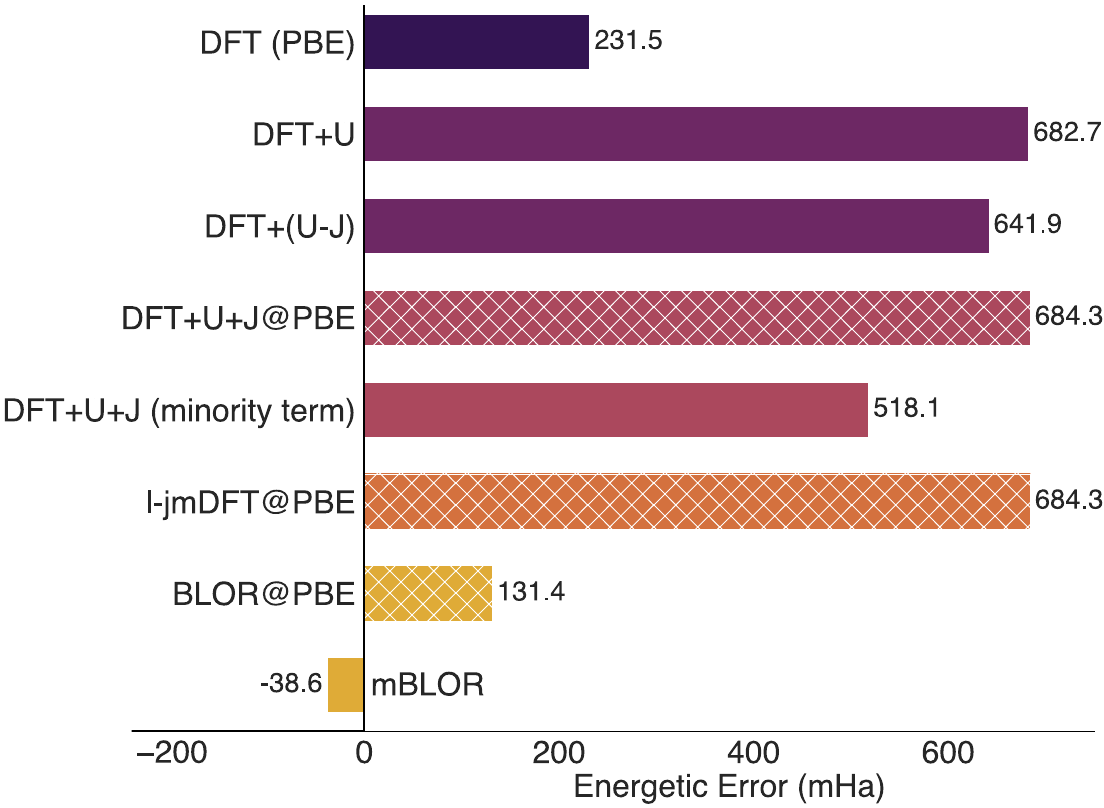}
\caption{  Bar chart of the errors in the total energy of singlet N$_2$ at an inter-nuclear separation length of 7$a_0$ using different corrective functionals  \cite{dudarevElectronenergylossSpectraStructural1998,himmetogluFirstprinciplesStudyElectronic2011,bajajCommunicationRecoveringFlatplane2017,bajajNonempiricalLowcostRecovery2019,burgessMathrmDFTTexttypeFunctional2023}. The hashing on the DFT+$U$+$J$, l-jmDFT and BLOR bars is used to indicate that these functionals were evaluated non-self consistently using the PBE density, as self-consistent application of these functionals (in a non-spin polarized DFT calculation), results in a symmetry broken ground state charge density. To avoid spurious spin-symmetry breaking, all other corrective functionals were evaluated self-consistently via a non-spin polarized DFT calculation. The raw DFT calculations were performed with the PBE functional \cite{perdewGeneralizedGradientApproximation1996}. The $U$ and $J$ parameters were evaluated using the simple $2 \times 2$ method~\cite{linscottRoleSpinCalculation2018} and $U^{\sigma}$ was computed via Eq.~\ref{eqn:Usigma}.  {All energetic errors are reported relative to twice the DFT(PBE) total energy of the symmetry broken nitrogen atom.}}
\label{fig:n2_bar_chart}
\end{figure}

\begin{figure}%[H]
\centering
\includegraphics[scale=0.2]{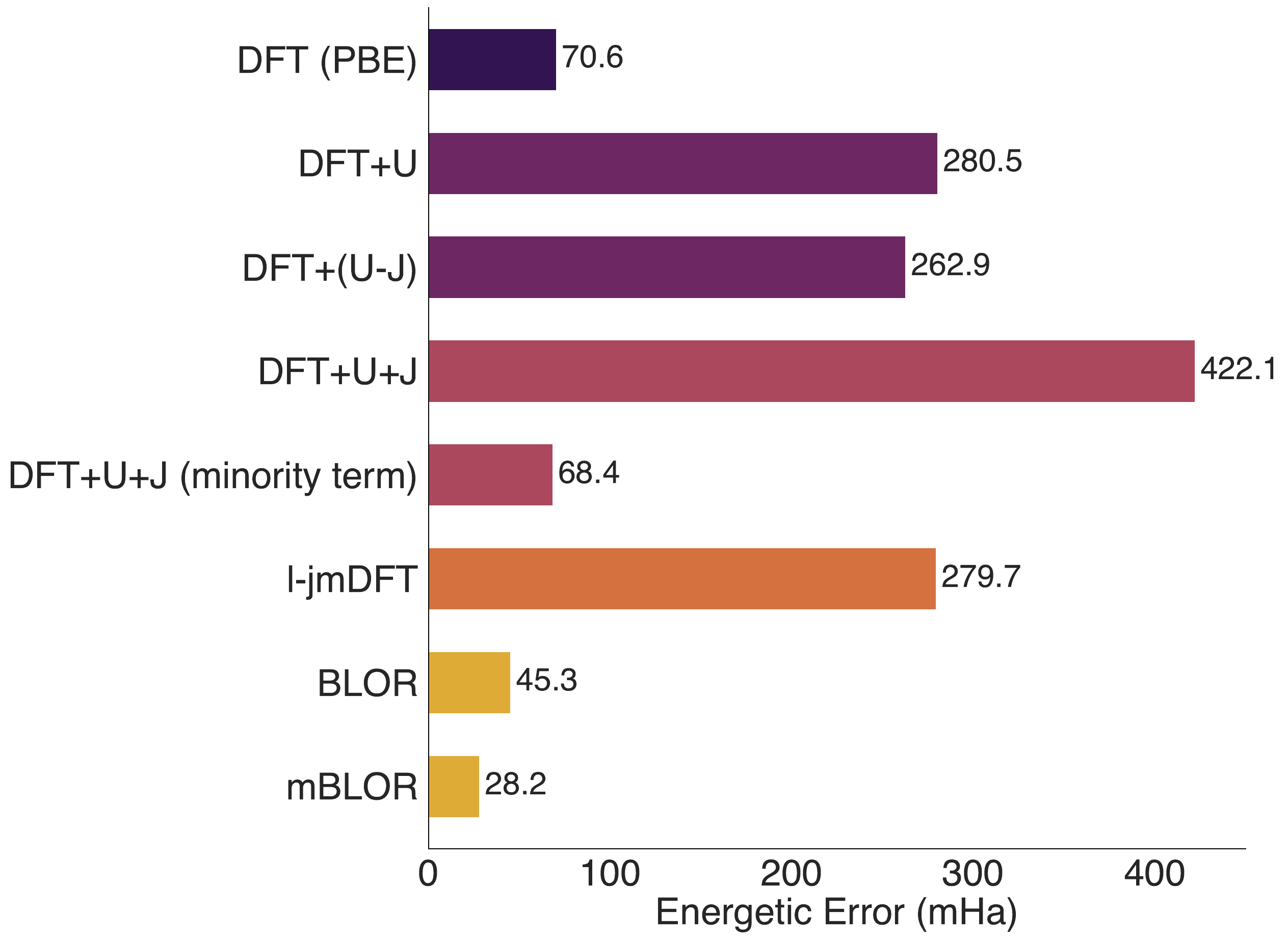}
\caption{  Bar chart of the errors in the total energy of singlet F$_2$ at an inter-nuclear separation length of 6$a_0$ using different corrective functionals  \cite{dudarevElectronenergylossSpectraStructural1998,himmetogluFirstprinciplesStudyElectronic2011,bajajCommunicationRecoveringFlatplane2017,bajajNonempiricalLowcostRecovery2019,burgessMathrmDFTTexttypeFunctional2023}. With the exception of the raw DFT(PBE) functional, all total energy errors have been evaluated self-consistently via a non-spin polarized DFT calculation to avoid spurious spin-symmetry breaking. The symmetry unbroken DFT(PBE) calculation failed to converge. The energy error for the DFT(PBE) functional has been evaluated by extrapolating the DFT$+U_{\rm in}$ total energy (for a series of values of $U_{\rm in}$), back to $U_{\rm in}=0$. The $U$ and $J$ parameters were evaluated using the simple $2 \times 2$ method~\cite{linscottRoleSpinCalculation2018} and $U^{\sigma}$ was computed via Eq.~\ref{eqn:Usigma}.  {All energetic errors are reported relative to twice the DFT(PBE) total energy of the symmetry broken fluorine atom.}}
\label{fig:f2_bar_chart}
\end{figure}

\begin{figure}%[H]
\centering
\includegraphics[scale=0.2]{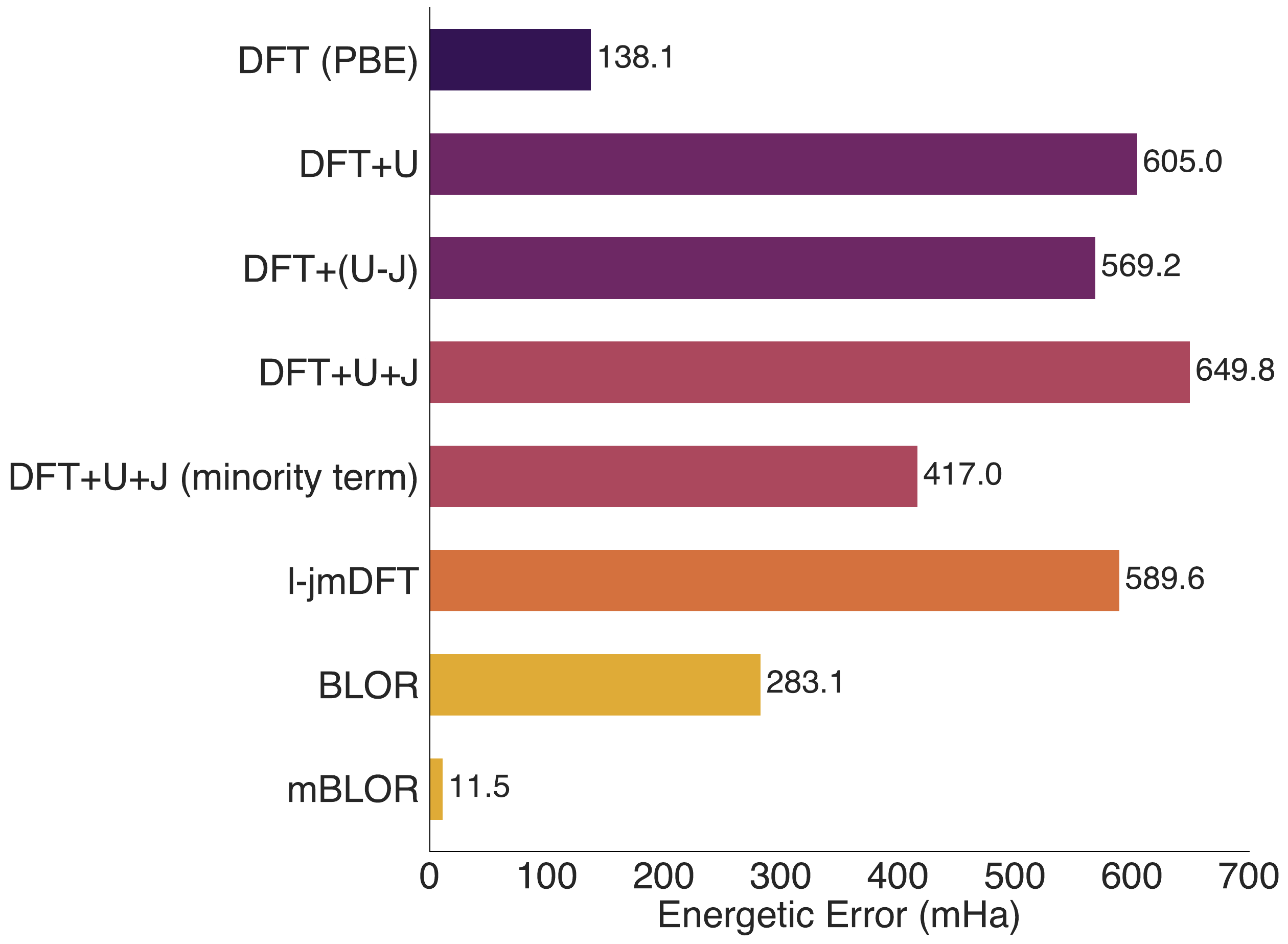}
\caption{ Bar chart of the errors in the total energy of non-spin polarized O$_2$ at an inter-nuclear separation length of 6$a_0$ using different corrective functionals  \cite{dudarevElectronenergylossSpectraStructural1998,himmetogluFirstprinciplesStudyElectronic2011,bajajCommunicationRecoveringFlatplane2017,bajajNonempiricalLowcostRecovery2019,burgessMathrmDFTTexttypeFunctional2023}. To simulate the non-spin polarized O$_2$ molecule, the spin up and spin down occupancies of the doubly degenerate  {GKS} orbitals at the Fermi level were all set equal to 0.5. The raw DFT calculations were performed with the PBE functional \cite{perdewGeneralizedGradientApproximation1996}. The $U$ and $J$ parameters were evaluated using the simple $2 \times 2$ method~\cite{linscottRoleSpinCalculation2018} and $U^{\sigma}$ was computed via Eq.~\ref{eqn:Usigma}. The ground state PBE total energy and spin resolved subspace occupancies were evaluated by linearly extrapolating the results from the O$_2$ calculations with a DFT+$U$+$J$-type stabilising potential back to $U_{\rm in}=J_{\rm in}=0$, see Appendix II for further details. Self-consistent application of the corrective functionals via non-spin polarized DFT calculations results in symmetry broken ground state charge densities, and so for this system the corrective functionals have been evaluated non-self consistently using the extrapolated PBE density.  {All energetic errors are reported relative to twice the DFT(PBE) total energy of the symmetry broken oxygen atom.}}
\label{fig:o2_bar_chart}
\end{figure}

The energy results are presented in Figs.~\ref{fig:n2_bar_chart}-\ref{fig:o2_bar_chart} for the stretched non-spin polarized N$_2$, O$_2$ and F$_2$ molecules, respectively. The DFT+$U$ and DFT$+(U-J)$ relative errors were computed using Dudarev et al.'s 1998 functional with the effective Hubbard parameter ($U_{\rm eff}$) set to $U$ and $U-J$, respectively. DFT+$U$+$J$ and DFT+$U$+$J$ (minority term) refers to the Hubbard corrective functional of Himmetoglu et al.~\cite{himmetogluFirstprinciplesStudyElectronic2011} excluding and including the minority spin term, respectively. The localized-jmDFT (l-jmDFT) functional refers to the localized subspace application corrective functional of Bajaj et al.~\cite{bajajCommunicationRecoveringFlatplane2017,bajajNonempiricalLowcostRecovery2019}, in which the use of
different functional forms for different flat-plane tiles was
pioneered. It  {is} applied in this work  using $U$ and $J$ parameters evaluated via the simple $2 \times 2$ method, which is not how the functional was originally intended to be applied.

Uncorrected approximate DFT, using the PBE approximation, yielded errors in the total energy as high as 232 mHa. Application of Dudarev's 1998 DFT$+U$ corrective functional significantly worsened these total energies, yielding errors as high as 682 mHa. This result suggests that Dudarev's DFT$+U$ functional, the most widely used Hubbard corrective functional in the literature, does not yield reliable total energies. By comparison, our newly developed mBLOR functional significantly improved  {upon} the raw DFT total energies, with relative errors below 39 mHa across all three test systems. 

 The original BLOR functional of Eq.~\ref{eqn:BLOR_trace_version} improved the total energies over raw DFT (PBE) in some cases but notably worsens the total energy compared to raw DFT for the non-spin polarized O$_2$ molecule, yielding an energy error of 283 mHa. The success of the mBLOR functional over the original BLOR functional suggests that inter-orbital interactions cannot be neglected in the development of DFT$+U$-type functionals that yield reliable total energies, although there may of course prove to be
 cases where BLOR outperforms mBLOR. The present bar charts show that only the mBLOR DFT$+U$ functional reduces the error in the raw DFT total energies across all three test systems.

\section{Tilted-plane based functional including  
inter-orbital corrections for many-body errors: mBLOR}
We are now ready to expand on our previous arguments and derive the mBLOR functional which can be applied to either spin polarized or non-spin polarized systems. The functional can also be used to enforce the global tilted plane condition on finite electronic systems that are in an external Zeeman field, or just 
when treated with spin-symmetry-broken approximate DFT. To derive this corrective functional we can no longer assume that the spurious curvature in the total subspace energy of an approximate XC functional with respect to spin up subspace occupancy $n^{\upharpoonright}$, is equal to the spurious curvature in the total subspace energy with respect to spin down subspace occupancy $n^{\downharpoonright}$. We denote these curvatures simply as $U^{\upharpoonright}$ and $U^{\downharpoonright}$, respectively. Now that $U^{\upharpoonright}\neq U^{\downharpoonright}$, the isosceles trapezoid shaped planes in the $E[N,M]$ energy surface of Fig.~\ref{fig:neon_energy_curve} will, in the simplest case, fracture into two triangular shaped planes. The breaking of this spin symmetry may be thought of as being due to an effective magnetic field $B_{\rm xc}({\bf r})$, acting on the subspace 
and due to the surrounding spin polarized material environment. This effective magnetic field is due to the differing spin resolved exchange-correlation potentials,  {and
given by} 
\begin{equation}
B_{\rm xc}({\bf r})=\frac{1}{2}\left(v_{\rm xc}^{\upharpoonright}({\bf r})-v_{\rm xc}^{\downharpoonright}({\bf r})\right),
\end{equation}
in atomic units.
A large variety of different $E[N, M] $ surfaces can arise due to this fracturing. In this study we consider only the simplest case, where each isosceles trapezoid shaped plane fractures into two triangular shaped planes, as presented in Appendix I. This fracturing pattern will occur if the subspace satisfies the following three criteria.
\begin{enumerate}[label=(\alph*)]
\item 
The total energy of the subspace obeys the strong convexity condition with respect to electron count,
\begin{equation}
2E[N,-M_j]< E[N-1,M_i]+E[N+1,M_k], 
\end{equation}
where $M_i$, $M_j$, and $M_j$ are the lowest energy magnetization states of the subspace with integer occupancies $N-1$, $N$ and $N+1$.
\item
The subspace-bath interaction energy varies linearly with spin-resolved occupancy.
\item  
Vertices in the subspace energy surface occur only when $M=\pm M_0$, where $M_0$ is the maximum magnetization of the subspace for a given integer value of occupancy. There exists $4{\rm Tr}[\hat{P}]$ such vertices for a given subspace.
\end{enumerate}
In order to enforce the localized tilted plane condition once on an entire multi-orbital subspace that meets criteria (a)-(c), the mBLOR functional is designed to satisfy four key conditions. It must:
\begin{enumerate}
    \item 
Be a continuous function of the subspace electron
count $N$ and subspace magnetization $M$.
    \item 
Yield no correction at the $4{\rm Tr}[\hat{P}]$ vertices. This is desirable because reasonable approximate functionals are expected to yield accurate total energies in this case.
    \item 
Have a constant curvature of $-U^{\sigma}$ with respect to $n^{\sigma}$. This is desirable because approximate functionals are expected to have a spurious curvature of $U^{\sigma}$ with respect to $n^{\sigma}$ due to their deviation from the localized flat plane condition, and this should be subtracted off.
    \item 
Have a constant curvature of $J$ with respect to $M$. This
is desirable because approximate functionals are expected to
have a spurious curvature of $-J$ with respect to $M$
(the minus is due to long-standing convention for defining $J$), 
and again  this should be subtracted off.
\end{enumerate}
The mBLOR functional is the mathematically unique functional that satisfies these four key conditions,  {and it is} given for each site by 
\begin{widetext}
\begin{align}
\label{eqn:BLOR++}
E_{\rm mBLOR}= \left\{
\begin{array}{*7{>{\displaystyle}c}}
%{\rm \hspace{8mm}}
\frac{U^{\upharpoonright}+U^{\downharpoonright}}{4}\left[{(N-N_0)}-{(N-N_0)}^2\right]
%{\rm \hspace{10.5mm}}
&+&
%{\rm \hspace{2mm}}
\frac{J}{2}\left[{M}^2-{N}^2\right]
%{\rm \hspace{5mm}}
&+&
{\frac{U^{\upharpoonright}-U^{\downharpoonright}}{4} F_{\rm AMSIE}^{\rm early}},%\left[N,M\right]},
%{\rm \hspace{6.75mm}}
&  \
&  \
%{\rm \hspace{5mm}}
N \leq {\rm Tr}[\hat{P}],
\\ \\
\frac{U^{\upharpoonright}+U^{\downharpoonright}}{4}\left[{(N-N_0)}-{(N-N_0)}^2\right]
%{\rm \hspace{2mm}}
&+&
%{\rm \hspace{2mm}}
\frac{J}{2}\left[{M}^2-{(N-2{\rm Tr}[\hat{P}])}^2\right]
%{\rm \hspace{5mm}}
%{\rm \hspace{2mm}}
&+&
\frac{U^{\upharpoonright}-U^{\downharpoonright}}{4} F_{\rm AMSIE}^{\rm late},%\left[N,M\right],
%{\rm \hspace{2mm}}
&  \
&  \
%{\rm \hspace{5mm}}
N > {\rm Tr}[\hat{P}],
\end{array}
\right.
\end{align}
which generates  the corresponding terms in the 
spin-dependent potential
\begin{align}
\label{eqn:BLORpotential++}
\hat{v}_{\rm mBLOR}^{\sigma}= \left\{
\begin{array}{*6{>{\displaystyle}c}}
%{\rm \hspace{8mm}}
\frac{U^{\upharpoonright}+U^{\downharpoonright}}{4}\left[1-2 \left(N - N_0 \right) \right]\hat{P}
%{\rm \hspace{10.5mm}}
&-&
%{\rm \hspace{10.5mm}}
2J N^{\bar{\sigma}}\hat{P}
%{\rm \hspace{6.75mm}}
&+&
%{\rm \hspace{2mm}}
{\frac{U^{\upharpoonright}-U^{\downharpoonright}}{4}\hat{v}^{\sigma \,{\rm early}}_{\rm AMSIE}[N,M],}
%{\rm \hspace{5mm}}
&  \
%{\rm \hspace{5mm}}
N \leq {\rm Tr}[\hat{P}],
 \\ \\
\frac{U^{\upharpoonright}+U^{\downharpoonright}}{4}\left[1-2\left(N - N_0 \right)\right]\hat{P}
%{\rm \hspace{2mm}}
&-&
%{\rm \hspace{2mm}}
2J\left[ N^{\bar{\sigma}}-{\rm Tr}[\hat{P}]\right]\hat{P}
%{\rm \hspace{2mm}}
&+&
%{\rm \hspace{2mm}}
{\frac{U^{\upharpoonright}-U^{\downharpoonright}}{4} \hat{v}^{\sigma \, {\rm late}}_{\rm AMSIE}[N,M],}
%{\rm \hspace{5mm}}
&  \
%{\rm \hspace{5mm}}
N > {\rm Tr}[\hat{P}].
\end{array}
\right.
\end{align}
\end{widetext}
Here, $N^{\bar{\sigma}}$ is the subspace  occupancy for the 
opposite spin channel to that indexed by $\sigma$.
The unitless quantity 
$F_{\rm AMSIE}\left[N,M\right]$ is here termed 
the asymmetric-many electron self interaction error function and $\hat{v}^{\sigma}_{\rm AMSIE}[N,M]$ is the correspondingly
generated spin-$\sigma$ asymmetric-MSIE potential operator. 
These are defined in Appendix I. The $F_{\rm AMSIE}\left[N,M\right]$ function arises whenever $U^{\upharpoonright}\neq U^{\downharpoonright}$. In such cases, a spin-symmetric and spin-asymmetric term is needed to properly mitigate the localized many electron self interaction error. 
 {A demonstration of how
functionals of this general type are
derived, and then proven to be unique,  
is provided in Supplementary Materials S-I 
and S-II of Ref.~\onlinecite{burgessMathrmDFTTexttypeFunctional2023}.}
The asymmetric-MSIE term has already been studied for s-valence systems by Burgess et al.~\cite{burgessMathrmDFTTexttypeFunctional2023} and its inclusion was found to be necessary for yielding accurate total energies for the stretched triplet H$_5^{+}$ ring system. The eight different forms of the asymmetric-MSIE function are presented in Appendix I along with the corresponding Asymmetric-MSIE potential operators.

It is worth emphasising that for non-spin polarized systems the mBLOR functional of Eq.~\ref{eqn:BLOR++} simplifies to the spin symmetric functional of Eq.~\ref{eqn:mBLORsym}. Furthermore, in the case of single orbital subspaces, Eq.~\ref{eqn:BLOR++} is equivalent to the original BLOR functional, which unlike standard DFT$+U$ functionals has been shown to yield accurate total energies for dissociated s-valence molecules, namely H$_2$, He$_2^+$, Li$_2$, Be$_2^+$ and triplet H$_5^{+}$.

It is worth mentioning that the implementation of the mBLOR
functional does not require substantial modifications to 
an existing simplified rotationally invariant 
DFT+$U$ code.
Terms in the energy
that ordinarily look like
$\rm{Tr} [ \hat{n}^\sigma \hat{n}^{\sigma'} ] $
are converted to the form 
$\rm{Tr} [ \hat{n}^\sigma ]
\rm{Tr} [ \hat{n}^{\sigma'} ] $, 
while terms in the potential that ordinarily look like 
$  \hat{n}^\sigma  $
are converted to the form 
$ N^\sigma  \hat{P} $.
If starting, as we have, with a code equipped with
simplified rotationally invariant 
DFT+U+J together which `$\alpha$' and `$\beta$'
potential shifts (that may of course 
be put to use other than for perturbation),
then  the aforementioned, along with the provision
for spin-dependent Hubbard U parameters, is all the in-code modification that is strictly needed for mBLOR.
The rest can be done by parameter rearrangement externally.
This is even more so the case for the
spin-symmetric approximation to BLOR, which requires no
code modification at all  {given
a rotationally-invariant DFT+U} 
starting point, as discussed
in the SM of 
Ref.~\onlinecite{burgessMathrmDFTTexttypeFunctional2023}.

\section{Spin-Asymmetric Molecular Test}
In the case of the singlet N$_2$, F$_2$ and non-spin polarized O$_2$ molecules, the mBLOR functional of Eq.~\ref{eqn:BLOR++} simplifies to the spin symmetric mBLOR functional, which has  been shown to yield energy errors below 39 mHa across these three test systems. The stretched Ne$_2^+$ molecule was selected as a spin-polarized test system for the mBLOR functional
due to its potential for harboring both symmetric and
asymmetric SIE in their many-body forms. Unlike the singlet N$_2$, F$_2$ and non-spin polarized O$_2$ molecules, at large inter-nuclear separation lengths a non-integer number of electrons 
(approximately 5.5 electrons) will localize on each of the two atomic sites  and hence the system will suffer from localized-MSIE. The atomic susbpaces of the Ne$_2^+$ molecule are maximally spin polarized and so the system is expected to suffer from negligible localized-SCE. 
Mori-S\'{a}nchez et al.~\cite{mori-sanchezDerivativeDiscontinuityExchange2014} have shown that standard DFT functionals either yield accurate total energies for the stretched H$_2^+$ molecule, which is dominated by localized MSIE, or accurate total energies for the stretched H$_2$ molecule, which is dominated by localized SCE, but never accurate total energies for both systems (of course some DFT functionals perform poorly for both). These four molecules at large separation lengths are therefore challenging test cases for any DFT$+U$-type functional, as in order to yield reliable total energies across all four test systems, the DFT$+U$ functional must be able to mitigate both forms of error.

\begin{figure}%[H]
\centering
\includegraphics[scale=0.2]{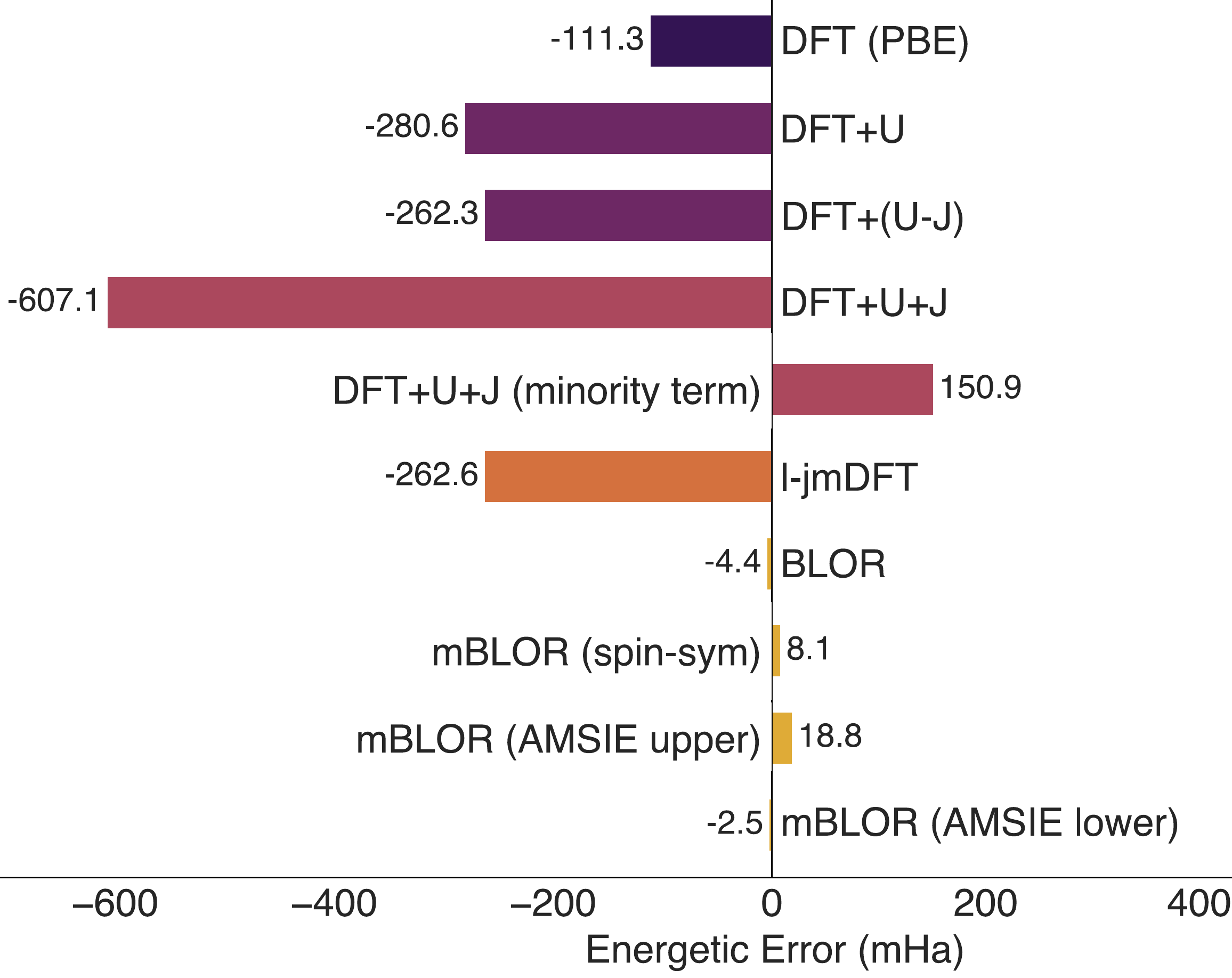}
\caption{ Bar chart of the errors in the total energy of the doublet Ne$_2^+$ molecule at an inter-nuclear separation length of 5$a_0$ using different corrective functionals  \cite{dudarevElectronenergylossSpectraStructural1998,himmetogluFirstprinciplesStudyElectronic2011,bajajCommunicationRecoveringFlatplane2017,bajajNonempiricalLowcostRecovery2019,burgessMathrmDFTTexttypeFunctional2023}. The raw DFT calculations were performed with the PBE exchange correlation functional \cite{perdewGeneralizedGradientApproximation1996}. The energy errors for the Hubbard type corrective functionals reported in this bar chart have been evaluated non-self consistently using the PBE density as all self consistent corrective functional calculations failed to converge for this system, with the notable exception of the mBLOR functional.  {All energetic errors are reported relative to the sum of the DFT(PBE) total energies of the neon atom and its spin-symmetry broken cation.}}
\label{fig:ne2+_bar_chart}
\end{figure}

In the dissociated limit the total energy of a cationic dimer X$_2^+$ should converge to the sum of the energies of the X atom and cation, in what might be considered the generalized 
extensivity condition 
\begin{equation}
\label{eqn:tot_energy_x2_cation}
E[{\rm X}_2^+]=E[{\rm X}]+E[{\rm X}^+].
\end{equation}
As was the case with the non-spin polarized test systems, we use the PBE total energies from the right-hand side of Eq.~\ref{eqn:tot_energy_x2_cation} as our reference value, i.e., the sum of the PBE X atom and X$^+$ cation total energies. The bar chart of the energy errors associated with each DFT$+U$-type functional are presented in Fig.~\ref{fig:ne2+_bar_chart}. 

As shown in the bar chart of Fig.~\ref{fig:ne2+_bar_chart}, the raw PBE functional  {yields} an energy error of 111 mHa and application of other DFT$+U$-type functionals significantly worsens this total energy, with the exception of BLOR and mBLOR. The  {appropriate} mBLOR functional reduces the PBE energy error to only -2.5 mHa. The original BLOR functional also performs exceptionally well for this test system, yielding an energy error of only -4.4 mHa. The success of the original BLOR functional in this particular test system is attributed to the fractional occupancy at the atomic site being limited to a single orbital, with the other two orbitals being almost fully occupied. 

When applying the mBLOR functional to spin polarized systems, as was the case for non-spin polarized systems, one must choose between the `early' and `late' versions of the mBLOR functional. In the case of the stretched Ne$_2^+$ molecule, the subspace is significantly more than half occupied $(N \approx 5.5)$ and thus the `late' version of the functional should be applied. For spin polarized systems, one must also choose the correct AMSIE function, of which there are 
eight. The eight different forms of the AMSIE function are given in Appendix I, along with a practical selection procedure to ensure the correct AMSIE function is chosen for a given subspace. In the case of the stretched Ne$_2^+$ molecule, the most important step in the AMSIE function selection procedure is choosing between the `lower' and `upper' versions of the AMSIE function. Both correctly yield no energy correction when the subspace occupancy is equal to that of the neutral neon atom, i.e., $N=6$ and $M=0$, but only the `lower' version of the AMSIE function yields no energy correction when the spin resolved subspace occupancy is equal to that of the spin polarized neon cation, i.e., $N=5$ and $M=\pm 1$. The `lower' version of the AMSIE function is thus the correct version to apply to this molecular system. Unsurprisingly, it yielded the smallest energy error of -2.5 mHa. If one instead choose (incorrectly) to apply the `upper' version of the AMSIE function, the energy error increases to 18.8 mHa  {(when also we incorrectly set 
$U^{\upharpoonright}>U^{\downharpoonright}$)}. Alternatively, choosing to omit the AMSIE function and instead apply the spin-symmetric mBLOR functional with $U^{\upharpoonright}=U^{\downharpoonright}=f^{\downharpoonright \downharpoonright}$, yields an energy error of 8.1 mHa.

The mBLOR functional is thus the only Hubbard type corrective functional that yielded low energy errors across all four test systems. The significantly improved total energies offered by the mBLOR corrective functional may facilitate the reliable prediction of chemical properties of transition metal compounds that have proven particularly challenging for current DFT$+U$-type functionals, such as spin-state energies~\cite{marianoBiasedSpinStateEnergetics2020,velaThermalSpinCrossover2020}, Heisenberg exchange coupling~\cite{macenultyOptimizationStrategiesDeveloped2023} and surface formation energies~\cite{zhaoStableSurfacesThat2019} but  {these} are  avenues for future research.

\section{Potentialized Explicit Derivative Discontinuities
for their convenient manifestation in the generalized Kohn-
Sham bandgap}
The mBLOR functional of Eq.~\ref{eqn:BLOR++} will exhibit derivative discontinuities~\cite{perdewPhysicalContentExact1983, shamDensityFunctionalTheoryEnergy1983, yangDerivativeDiscontinuityBandgap2012,  mori-sanchezDerivativeDiscontinuityExchange2014, gouldKohnShamPotentialsExact2014,kraislerKohnShamManyElectron2021,cernaticEnsembleDensityFunctional2021} at integer values of subspace occupancy due to the functional's explicit dependence on $N$. Therefore, it is neither a differentiable functional of the electron density or the first-order non-interacting density matrix. The mBLOR functional thus fits into category D of Yang et al.'s~\cite{yangDerivativeDiscontinuityBandgap2012} functional classification scheme, which is defined as any ``{\it functional of the density or first-order density matrix with explicit discontinuity.}" This derivative discontinuity $\Delta_{\rm xc}^{N}$, will contribute to the quasi-particle bandgap, $\Delta$, whenever increasing the global electron count $N_{\rm tot}$ by one causes the subspace occupancy at any of the atomic sites to increase through an integer value,  so that $\lfloor N \rfloor =N_0$ also increases by one. This contribution to the quasi-particle bandgap $\Delta_{\rm xc}^{N}$, will not appear in the Kohn-Sham gap $\Delta_{\rm KS} $ or the Generalized Kohn-Sham gap $\Delta_{\rm GKS}$, unless something is done.
This is unlike the situation in DFT+$U$
functionals such as the Dudarev simplified rotationally-invariant 
one, as those comprise  {only} \emph{implicit}
derivative discontinuities via their subspace
projected density-matrix dependence
 {(those which give the
gap opening of order  $U-J$ in GKS).}
An implicit derivative discontinuity may arise in mBLOR due to 
possibly different subspace projection weightings at
the valence (occupied) and conduction (virtual) band edges, 
and this is expected to be a small effect
for most applications of interest.
However, substantial \emph{explicit}
derivative discontinuities appear in mBLOR,  
due to both the $N_0$ terms and to
switch between early and late version, and these are rather unusual
in the approximate or corrective functional contexts.

Ordinarily, calculating the mBLOR fundamental gap
would not be strictly possible 
only via the single-particle eigenspectrum, because
explicit derivative discontinuities do not manifest there, 
but it would instead require adding a separate contribution 
to the  bandgap post hoc, as in 
\begin{equation}
\Delta=\Delta_{\rm GKS}+\Delta_{\rm xc}^{N}.
\end{equation}
Even this seems difficult as, opposed to being a constant global derivative discontinuity, the derivative discontinuity of the mBLOR functional is a non-local
operator
in character (non-local in the sense of exhibiting more than
one spatial argument) and more generally a sum over sites
of those. For a single site, we may write
\begin{equation}
\label{eqn:non_local_derivative_discontinuity}
\hat{\Delta}_{\rm xc}^{N}=\widetilde{\Delta}_{\rm xc}\left[N\right]\hat{P}.
\end{equation}
The $\widetilde{\Delta}_{\rm xc}$ function of Eq.~\ref{eqn:non_local_derivative_discontinuity} itself depends on the subspace occupancy $N$. For subspaces with $U^{\upharpoonright}\geq U^{\downharpoonright}$, we have 
\begin{align}
\label{eqn:derivative_discontinuity_blor}
\widetilde{\Delta}_{\rm xc} = \left\{
\begin{array}{*1{>{\displaystyle}c}}
U^{\downharpoonright}-\frac{U^{\upharpoonright}-U^{\downharpoonright}}{2}N_0, \qquad N<{\rm Tr}[\hat{P}],  \\ \\
U^{\upharpoonright} +2J{\rm Tr}[\hat{P}]-\frac{U^{\upharpoonright}-U^{\downharpoonright}}{2}{\rm Tr}[\hat{P}],\quad N={\rm Tr}[\hat{P}],  \\ \\
U^{\upharpoonright}-\frac{U^{\upharpoonright}-U^{\downharpoonright}}{2}\left(2{\rm Tr}[\hat{P}]-N_0\right),\enspace N>{\rm Tr}[\hat{P}],
\end{array}
\right.
\end{align}
 {when $N$ crosses integer values, 
at which we use $N_0=  N-1$.} For non-spin polarized systems $\widetilde{\Delta}_{\rm xc}=U^{\sigma}=U-J$, except at half filling. The apparently larger than might  {be} expected 
 contribution of $\widetilde{\Delta}_{\rm xc}$ to the bandgap at subspace half filling
 $(N={\rm Tr}[\hat{P}])$ is explained  in the Bandgap Analysis section. 
%\comment{~\cite{zhaoGlobalLocalCurvature2016}}
 {For concision, we do not consider 
here the more involved forms
of $\widetilde{\Delta}_{\rm xc} $
that arise when moving from `lower' to `upper' type
tiles at fixed $N_0$.}

Accounting for the $\Delta_{\rm xc}^{N}$ contribution to the quasi-particle band-gap post calculation would be a rather 
disappointing if necesary, given that one of the most 
prominent and convenient features of conventional DFT+$U$  
is that its derivative discontinuity manifests automatically
in the Generalized Kohn-Sham bandgap.
To solve this and restore
the convenient bandgap opening in the 
GKS eigensystem, we have found that it is possible to 
`potentialize' the mBLOR explicit derivative discontinuity, 
that is to represent  {it as} an additional term in the potential. 
As the derivative discontinuity arises when adding an 
additional electron to the neutral system, only the conduction
band should be affected by this potential.
The additional term that achieves a shift of $\widetilde{\Delta}_{\rm xc}\left[N\right]$ to the portion of the conduction band that projects onto the localized subspace is
\begin{equation}
\label{eqn:additional_potential}
\hat{v}^{\sigma}_{\Delta \rm{xc}}
=\widetilde{\Delta}_{\rm xc}\left[N\right](1-\hat{\rho})\hat{P}(1-\hat{\rho}),
\end{equation}
The corresponding energy term may be found, by integration, 
and it is 
\begin{align}
\label{eqn:additional_energy_term}
E_{\Delta \rm{xc}} =&{}\frac{1}{2}\widetilde{\Delta}_{\rm xc}\left[N\right] {\rm Tr} \bigg[  \left(\hat{\rho} - \frac{\hat{\rho}^2}{2}\right)  \hat{P} \left(1-\hat{\rho}\right)  \nonumber  \\ &{}+  \left(1-\hat{\rho}\right) \hat{P} \left(\hat{\rho} - \frac{\hat{\rho}^2}{2}\right)\bigg] \\
=&{}\frac{1}{2}\widetilde{\Delta}_{\rm xc}\left[N\right] {\rm Tr} \bigg[  \hat{P} \left(2 \hat{\rho} - 3  \hat{\rho}^2 +   \hat{\rho}^3 \right)    
\nonumber \bigg].
\end{align}
This yields a vanishing correction to the total energy  under the assumption that $\hat{\rho}=\hat{\rho}^2$, which holds for insulators
at zero temperature as well as for molecules with non-degenerate ground states. In practice, we have found as long as the initial guess
for the electronic structure exhibts a gap, it is possible to 
apply the $\hat{v}^{\sigma}_{\Delta \rm{xc}}$ self-consistently
without adding the vanishing $E_{\Delta \rm{xc}}$ 
to $E_{\rm{mBLOR}}$ at
all, without appreciable loss of convergence performance.
Of course, for finite-temperature systems, metals, or
degenerate molecules, it would be necessary to add 
$E_{\Delta \rm{xc}}$ to the energy, for consistency.

%Our complete corrective functional is given by the combination of Eqs.~\ref{eqn:BLOR++},~\ref{eqn:derivative_discontinuity_blor} \& ~\ref{eqn:additional_energy_term},
%\begin{widetext}
%\begin{align}
%\label{eqn:BLOR_final}
%E_{\rm mBLOR}= \left\{
%\begin{array}{*1{>{\displaystyle}c}}
%\frac{U^{\upharpoonright}+U^{\downharpoonright}}{4}\left[{(N-N_0)}-%{(N-N_0)}^2\right%]
%+ \frac{J}{2}\lef%t[{M}^2-{N}^2\right]
%+
%{\frac{U^{\upharpoonright}-U^{\downharpoonright}}{4}F_{\rm %AMSIE}^{\rm early}\left[N,M\right]}
%+
%E_{\rm add},
%\quad
%N \leq {\rm Tr}[\hat{P}], 
%\\   \\ 
%\frac{U^{\upharpoonright}+U^{\downharpoonright}}{4}\left[{(N-N_0)}-%{(N-N_0)}^2\right]
%+  \frac{J}{2}\left[{M}^2-{(N-2{\rm Tr}[\hat{P}])}^2\right]
%+ {\frac{U^{\upharpoonright}-U^{\downharpoonright}}{4}F_{\rm %AMSIE}^{\rm late}\left[N,M\right]}
%+
%E_{\rm add},
%\quad
%N > {\rm Tr}[\hat{P}].
%\end{array}
%\right.
%\end{align}
%\end{widetext}

\section{Bandgap  Analysis}
To illustrate the corrective nature of the mBLOR functional, 
we may consider its application to a homo-nuclear spin-polarized system of transition metal atoms at half occupancy, so that each atomic d-orbital subspace is occupied by five spin up electrons and zero spin down electrons. For simplicity assume that the total subspace occupancy is infinitesimally greater than five and $U^{\upharpoonright}=U^{\downharpoonright}=U$, so that the late version of the mBLOR functional is applied and the $F_{\rm AMSIE}^{\rm late}\left[N,M\right]$ function can be neglected. Furthermore, assume that the highest occupied and lowest unoccupied  {GKS} orbitals project perfectly onto the atomic subspace. The spin-up subspace is maximally occupied so that the highest occupied GKS eigenvalue $(\epsilon_{\rm HOMO}$) is spin-up and the lowest unoccupied GKS eigenvalue $(\epsilon_{\rm LUMO}$) is spin-down. 

In this particular example, the subspace is at a vertex in the $E [N, M ]$ surface and so, to first-order in perturbation theory, the mBLOR corrective functional offers no correction to the total energy. If we now turn our attention to the potential. 
Excluding the $\widetilde{\Delta}_{\rm xc}\left[N\right]$ term, application of the mBLOR corrective functional will, to first order perturbation theory, shift the highest occupied GKS eigenvalue by
\begin{equation}
\Delta \epsilon_{\rm HOMO} =\frac{U}{2}+10J , 
\end{equation}
and shift the lowest unoccupied GKS eigenvalue by
\begin{equation}
\Delta \epsilon_{\rm LUMO} =\frac{U}{2}.
\end{equation}
This results in a closing of the GKS gap by  $10J$,
a very surprising result at first.
However, the 
incorporation  {of the} $\widetilde{\Delta}_{\rm xc}\left[N\right]$ term will counteract this affect precisely, by opening the gap by order $U+10J$, resulting in an effective gap opening of $U$. These results are summarized in Table~\ref{table:BLOR-gap-opening}.

\begin{table}
\centering
\begin{tabular}{|c|c|c|c|} 
 \hline
 $\Delta \epsilon_{\rm LUMO}$ & $\Delta \epsilon_{\rm HOMO}$ & $\widetilde{\Delta}_{\rm xc}\left[N\right]$ & $\Delta$
 \\ 
   \hline
   
   $\frac{1}{2}U$ & $\frac{1}{2}U+10J$ & $U+10J$ & $U$
 \\ 

 \hline  
\end{tabular}
\caption{A summary of the effect of the mBLOR corrective functional on a homo-nuclear spin polarized d-orbital system at half occupancy to first-order perturbation theory. $\Delta \epsilon_{\rm LUMO}$ and $\Delta \epsilon_{\rm HOMO}$ denote the shift to the lowest unoccupied and highest occupied GKS eigenvalues excluding the affect of the $\widetilde{\Delta}_{\rm xc}\left[N\right]$ term. The third column presents the gap opening due to the $\widetilde{\Delta}_{\rm xc}\left[N\right]$ term and finally $\Delta$ denotes the mBLOR functional's total contribution to the GKS gap.}
\label{table:BLOR-gap-opening}
\end{table}

\begin{figure*}
  \includegraphics[width=\textwidth]{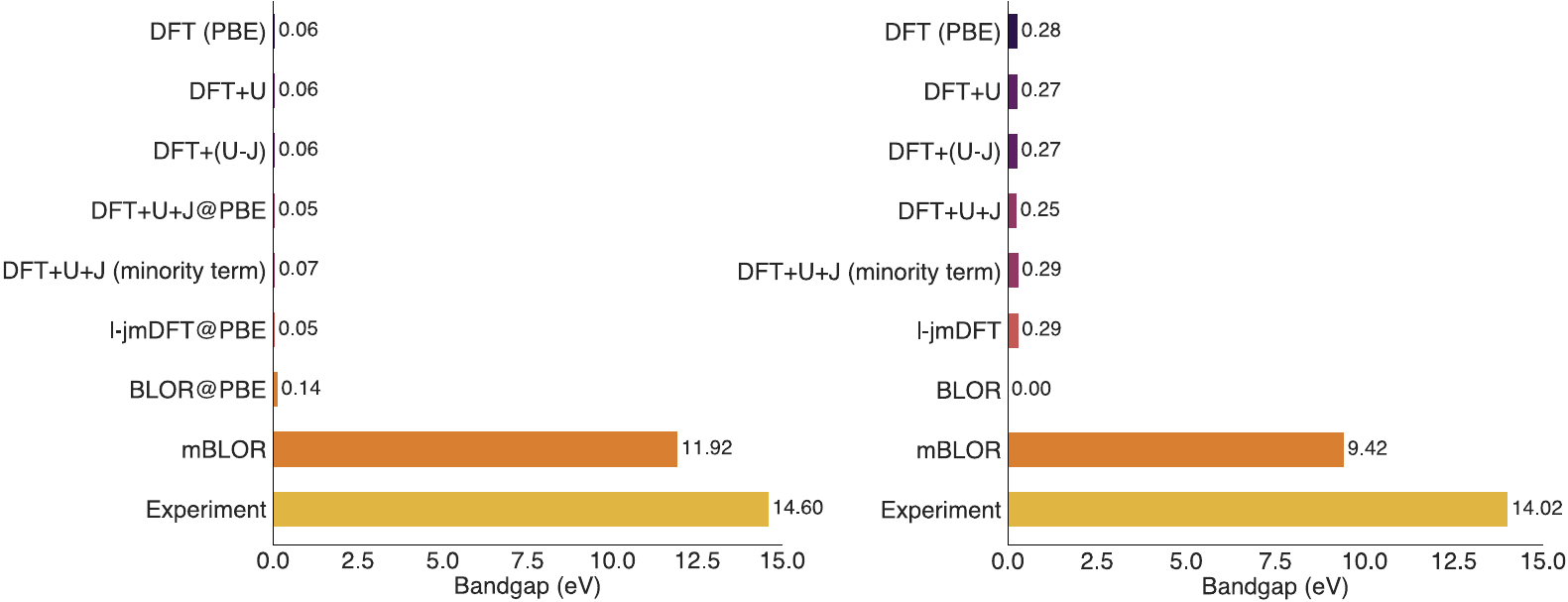}
\caption{  Bar chart of the predicted bandgaps of the stretched singlet N$_2$ and F$_2$ molecules at an inter-nuclear separation length of 7$a_0$ and 6$a_0$, respectively, using different corrective functionals \cite{dudarevElectronenergylossSpectraStructural1998,himmetogluFirstprinciplesStudyElectronic2011,bajajCommunicationRecoveringFlatplane2017,bajajNonempiricalLowcostRecovery2019,burgessMathrmDFTTexttypeFunctional2023}. The PBE exchange correlation functional \cite{perdewGeneralizedGradientApproximation1996} was employed through out. With the exception of the mBLOR functional, all bandgap predictions needed to be evaluated via non-spin polarized DFT calculations to avoid spurious spin symmetry breaking. For the N$_2$ molecule, the DFT+$U$+$J$, l-jmDFT and BLOR functionals were evaluated non-self consistently using the PBE density, as self-consistent application of these functionals (in a non-spin polarized DFT calculation), results in a symmetry broken ground state charge density.  For the F$_2$ molecule, the symmetry unbroken DFT(PBE) calculation failed to converge and so the bandgap for the DFT(PBE) functional has been evaluated by extrapolating the DFT$+U_{\rm in}$ bandgap (for a series of values of $U_{\rm in}$), back to $U_{\rm in}=0$. The reference experimental bandgaps for the stretched N$_2$ and F$_2$ molecules were assumed to be equal to their respective bandgaps at the dissociated limit and as such were evaluated as the difference between the reference atomic ionization potential~\cite{NIST_ASD} and atomic electron affinity~\cite{rienstra-kiracofeAtomicMolecularElectron2002} of the respective elements. One should note that the reference value for the electron affinity of the nitrogen atom is negative at -0.07 eV.}
\label{fig:f2_bandgap_bar_chart}
\end{figure*}

While further testing will be needed,
based on the systems tested here, the mBLOR functional may  {offer}  improved bandgaps over other DFT$+U$-type functionals. In the case of the stretched N$_2$ and F$_2$ molecules, the mBLOR functional is found to be the only known DFT$+U$ functional to open the bandgap
appreciabily. In these systems an approximately integer number of electrons localizes on each atomic site, so we employ the derivative discontinuity correction term when applying the mBLOR functional. In the stretched N$_2$ system the subspace occupancy is approximately equal to 3.0 and so we let $\widetilde{\Delta}_{\rm xc}\left[N\right]=U^{\upharpoonright}+6J$ and in the stretched F$_2$ system the subspace occupancy is approximately equal to 5.0 and so we let $\widetilde{\Delta}_{\rm xc}\left[N\right]=U^{\upharpoonright}$, where we recall that $U^{\upharpoonright}=U^{\downharpoonright}$ in these spin symmetric systems. The raw DFT calculation at the PBE level yields a gap of only 0.06 eV and 0.28 eV, respectively,
 {for N$_2$ and F$_2$}. Application of other DFT$+U$-type functionals including Dudarev's 1998 Hubbard functional and the original BLOR functional fail to improve the poor PBE result. In contrast, the mBLOR functional opens the gap to 11.92 eV and 9.42 eV, respectively. These bandgaps are still smaller than the anticipated values of 14.60 and 14.02 eV, but this represents an improvement by over one-order of magnitude compared to all other DFT$+U$ functionals tested here. The anticipated values are based on the assumption that the fundamental bandgap of the stretched N$_2$ and F$_2$ molecules are at the dissociated limit, in which case the gap will be equal to that of the isolated nitrogen or fluorine atom, 
respectively. These results are presented in Fig~\ref{fig:f2_bandgap_bar_chart}.

\begin{figure}%[H]
\centering
\includegraphics[scale=0.2]{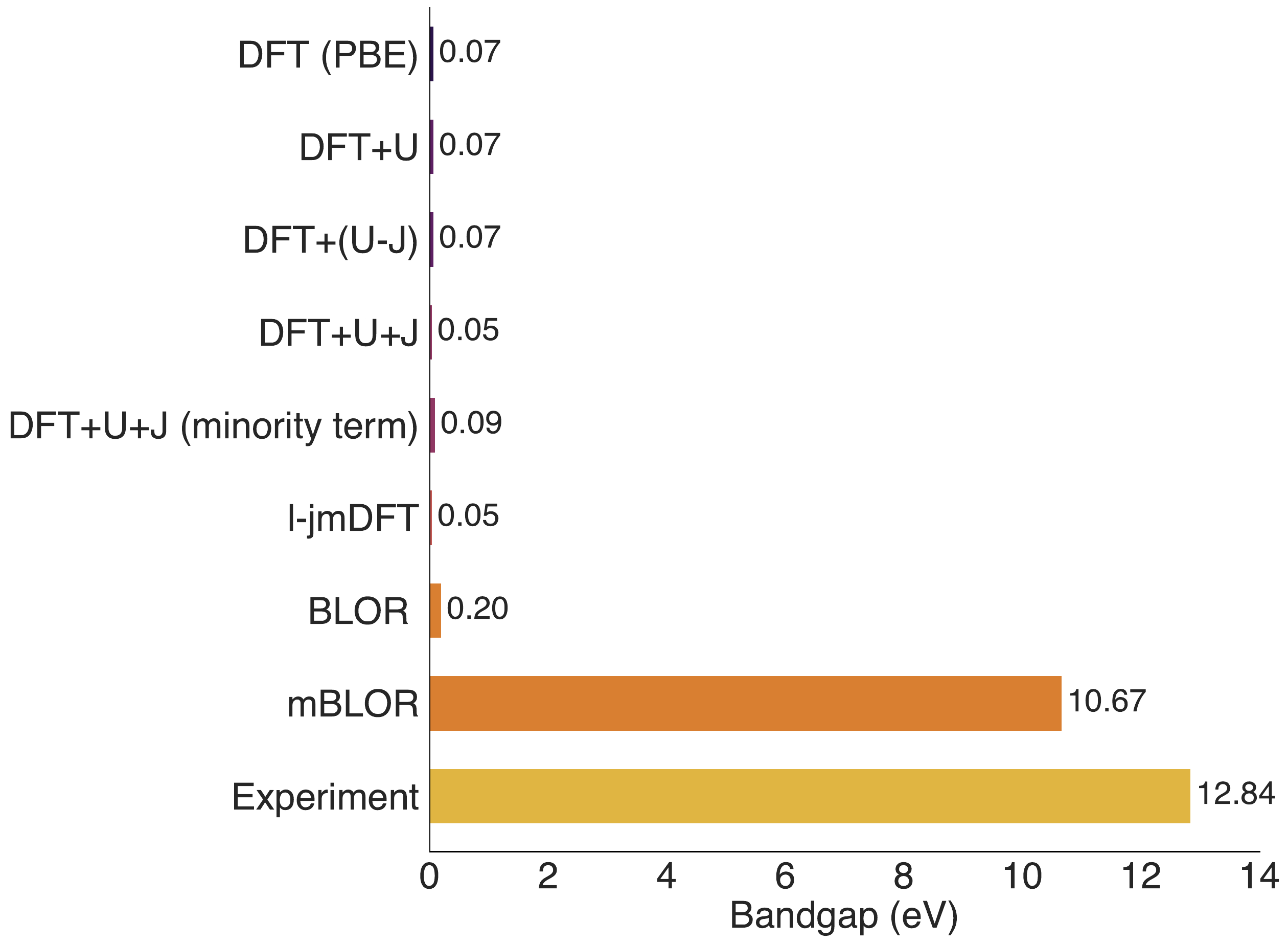}
\caption{ Bar chart of the predicted bandgaps of the stretched singlet H$_2$ molecule at an inter-nuclear separation length of 9$a_0$ using different corrective functionals  \cite{dudarevElectronenergylossSpectraStructural1998,himmetogluFirstprinciplesStudyElectronic2011,bajajCommunicationRecoveringFlatplane2017,bajajNonempiricalLowcostRecovery2019,burgessMathrmDFTTexttypeFunctional2023}. The raw DFT calculations were performed with the PBE exchange correlation functional \cite{perdewGeneralizedGradientApproximation1996}. The DFT+ $U$ and DFT$+(U-J)$ relative errors were computed using Dudarev et al.'s 1998 functional with the effective Hubbard parameter ($U_{\rm eff}$) set as $U$ and $U-J$, respectively. The reference experimental bandgap for the stretched H$_2$ molecule was assumed to be equal to its respective bandgap at the dissociated limit and, as such, was evaluated as the difference between the reference atomic ionization potential~\cite{NIST_ASD} and atomic electron affinity~\cite{rienstra-kiracofeAtomicMolecularElectron2002}. All bandgap predictions were evaluated via spin-polarized DFT calculations, for this system all corrective functionals preserved spin symmetry.}
\label{fig:h2_bandgaps_bar_chart}
\end{figure}

The failure of Dudarev's functional to improve the bandgap prediction for the stretched, symmetry unbroken N$_2$ and F$_2$ molecules compared to raw DFT is not a result unique to these two molecules, in fact Dudarev's functional will fail to open the bandgap of any symmetry unbroken, stretched, neutral, homo-nuclear dimer. This can be readily explained using the stretched H$_2$ molecule. At these large inter-nuclear separation lengths, the highest occupied and lowest unoccupied  {GKS} orbitals  can be approximated as a linear combination of the hydrogenic atomic orbitals centred at the atomic sites $1$ and $2$,
\begin{equation}
    \ket{\psi_{\rm{ {GKS}}}}=\frac{1}{\sqrt{2}}\left(\ket{\psi_1}\pm \ket{\psi_2}\right),
\end{equation}
and the spin resolved  {GKS} 
density operator is 
\begin{equation}
\label{h2-spin-density-operator}
\hat{\rho}^{\sigma}=\frac{1}{2}\left(\ket{\psi_1}\bra{\psi_1}+ \ket{\psi_1}\bra{\psi_2}+\ket{\psi_2}\bra{\psi_1}+\ket{\psi_2}\bra{\psi_2}\right).
\end{equation}
Dudarev's functional contributes an additional term to the 
 {GKS} potential,
\begin{equation}
    \hat{v}_{\rm u}=\frac{U_{\rm eff}}{2}\left(\sum_I\hat{P}_I-2\hat{P}_I\hat{\rho}^{\sigma}\hat{P}_I\right),
\end{equation}
where the summation runs over the two atomic sites and $\hat{P}_I$ is the atomic projection operator centred at site $I$, which in the case of an s-valence system, will be composed of a single atomic orbital
\begin{equation}
\label{h2-atomic-projector}
\hat{P}_I=\ket{\psi_I}\bra{\psi_I}.
\end{equation}
To first order perturbation theory, Dudarev's DFT$+U$ correction to the highest occupied  {GKS eigenvalue is} 
\begin{align}
&\braket{\psi_{\rm HOGKS}|\hat{v}_{\rm u}|\psi_{\rm HOGKS}}= \nonumber \\ & \frac{U_{\rm eff}}{2}\sum_{IK}\frac{1}{\sqrt{2}}\braket{\psi_I|\left(\sum_J\hat{P}_J-2\hat{P}_J\hat{\rho}^{\sigma}\hat{P}_J\right)\frac{1}{\sqrt{2}}|\psi_K},
\end{align}
which through use of Eqs~\ref{h2-spin-density-operator} \& \ref{h2-atomic-projector} one can readily show is equal to zero assuming the H$_2$ molecule is sufficiently stretched so that $\braket{\psi_I|\psi_J}\approx 0$, when $I \neq J$. Therefore, to first order perturbation theory, Dudarev's functional will offer no correction to the highest occupied  {GKS} eigenvalue. The same is true for the lowest unoccupied  {GKS} eigenvalue and thus Dudarev's functional cannot be used to open the  {GKS} gap of the stretched H$_2$ molecule. The same argument can also be applied to any symmetry unbroken, stretched, neutral, homo-nuclear p-block and d-block dimer (we exclude  van der Waals molecules in this discussion), assuming the  {GKS} orbitals are quasi-degenerate at the raw DFT level and can be approximated as an equally weighted linear combination of one p/d orbital centred on each atomic site. To further highlight the failure of conventional DFT+U functional to improve the bandgap of such systems without spurious symmetry breaking, we also present the bandgap of the symmetry unbroken stretched H$_2$ molecule in Fig~\ref{fig:h2_bandgaps_bar_chart}. By comparison, the mBLOR functional yields a bandgap of 10.67 eV, in close but not perfect agreement with the anticipated value of 12.84 eV.

For the stretched Ne$_2^+$ molecule, a non-integer number of electrons localizes on each atomic site ($N \approx 5.5$) and so in this case no derivative discontinuity correction will be applied with the mBLOR functional. The fundamental bandgap of the Ne$_2^+$ molecule is equal to the difference in magnitude of the ionization potential and electron affinity,
\begin{equation}
\label{eqn:ne2plus_bandgap}
\Delta=\left(E[{\rm Ne}_2^+]-E[{\rm Ne}_2^{+2}]\right)-\left(E[{\rm Ne}_2]-E[{\rm Ne}_2^+]\right).
\end{equation}
In the dissociated limit, each total energy term in Eq~\ref{eqn:ne2plus_bandgap} simplifies to the sum of two isolated neon atom/ion energies,
\begin{align}
\Delta=& \left(E[{\rm Ne}]+E[{\rm Ne}^+]- 2E[{\rm Ne}^+]\right)\nonumber \\ & -\left(2E[{\rm Ne}]-E[{\rm Ne}]-E[{\rm Ne}^+]\right)=0.
\end{align}
Therefore the anticipated value of the bandgap of the stretched Ne$_2^+$ is 0 eV, assuming the fundamental bandgap of the Ne$_2^+$ molecule at 5$a_0$ is at the dissociated limit. The raw DFT calculation at the PBE level yields a bandgap of 0.327 eV and similarly, the mBLOR functional yields a gap of 0.359 eV. These small residual gaps can be attributed to the Ne$_2^+$ molecule not completely reaching its dissociated limit at this inter-nuclear separation length. 

 {\section{Subspace Population Analysis}}
 {By virtue of the mBLOR functional being an explicit functional of the total subspace occupancy $N$ and magnetization $M$, as opposed to the spin resolved orbital occupancies $n^{\sigma}_{\rm mm'}$, mBLOR will typically be more sensitive to the choice of projection operator $\hat{P}$ used to define the localized subspace. Very
simplistically, we expect the sensitivity of the 
energy for each subspace to  approximately  
be an factor of $\rm{Tr}[ \hat{P}]$
times larger than that of BLOR or Dudarev's functional.}

 {In the case of the highly stretched molecular test systems employed in this study, the atomic orbitals represent an near-ideal choice for the subspace projection operator $\hat{P}$. However, due to the sensitivity of the mBLOR functional to the subspace population analysis, e.g., to spillage and
overlap, future application of this corrective functional to standard molecular and solid state systems in the bonding regime will require careful analysis of the suitability of using the atomic orbitals to define the subspace projection operator, as is standard practice for use with traditional DFT$+U$-type functionals. The combination of mBLOR with projectors
of Wannier-function type is a very attractive avenue
for future exploration.}

 {\section{Future Work}}
 {Future studies are expected to focus on the application of the mBLOR corrective functional to strongly correlated 3d transition metal oxides. For such solid state systems, particular emphasis will need to given to the mBLOR potential $\hat{v}_{\rm mBLOR}^{\sigma}$. The relatively large correction offered by the mBLOR potential could significantly alter the electronic structure of the system compared to that of the bare DFT calculation. In particular, the symmetric-MSIE term will drive the subspace towards integer occupancy.} 

 {As presented in Eq.~\ref{eqn:BLORpotential++}, the mBLOR potential is defined piecewise with respect to subspace occupancy $N$, due to its dependence on $N_0=\lfloor N \rfloor$ and due to the existence of both an early and late version of the functional. This piecewise definition could present convergence issues, particularly if the subspace is driven to integer occupancy, as in such cases a small update in the value of $N$ could change the value of $N_0=\lfloor N \rfloor$, which will result in a large change to the corrective potential.}

 {The performance of the derivative discontinuity correction term $\hat{v}^{\sigma}_{\Delta \rm{xc}}$, when applied to transition metal oxide systems will also need a thorough investigation. In general, the use of this term is advisable, 
as it will typically be needed to keep the GKS gap open.
Due to the presence of the subspace projection operator $\hat{P}$ in Eq.~\ref{eqn:additional_potential}, the derivative discontinuity correction will shift only the portion of the conduction band that projects onto the localized subspace. However, the conduction band edge of transition metal oxide systems is often of mixed atomic character and therefore the correct treatment of the bandgap may demand application of the derivative discontinuity correction to, e.g., the oxygen 2p and transition metal 4s atomic subspaces, in addition to the transition metal 3d subspace where the DFT$+U$ correction is more traditionally applied.}

 {
Further investigation and likely 
refinement of the mBLOR functional will thus  be necessary in 
order for the method to enjoy large-scale adoption. 
If the promising energetic results obtained for the stretched molecular test systems can be repeated for transition metal oxides, this novel DFT$+U$ functional could offer a reliable yet computationally inexpensive technique of simulating a suite of physical and chemical properties of transition metal oxide systems that depend on the total energy, such as redox potentials in Li-ion battery cathodes~\cite{shishkinChallengesComputationalEvaluation2017,muellerEvaluationTavoriteStructuredCathode2011,cococcioniEnergeticsCathodeVoltages2019}, surface formation energies for photo-catalysts~\cite{zhaoStableSurfacesThat2019,torrellesGeometricStructure$mathrmTiO_20112ifmmodetimeselsetexttimesfi1$2008,batzillFundamentalAspectsSurface2011} and crystallographic structures of high $T_{\rm c}$ superconductors in their normal state~\cite{beenElectronicStructureTrends2021,furnessAccurateFirstprinciplesTreatment2018}. Many challenges and opportunities 
thus remain in the emerging area of expediently correcting
approximate DFT \emph{in situ} 
using double-counting free generalized Hubbard corrections
such as BLOR and mBLOR.
The potential intellectual and
practical rewards are considerable.}

\section{Conclusions}
A double-counting approximation free DFT$+U$-type corrective functional named mBLOR has been derived from first principles to enforce the tilted plane condition on localized, multi-orbital subspaces. This  corrective functional is designed to depend only on the total subspace occupancy and magnetization, 
so that the many-body (inter-orbital including) self-interaction
(symmetric and asymmetric) and static correlation errors
are addressed in a very cost-effective and easy
to implement way. 
This approach ensures consistency between how subspace-resolved errors are measured, via the Hubbard and Hund parameters,
and how they are corrected. The formalism is readily
applicable to orbital-free DFT, ensemble DFT, 
or many other types of electronic structure theory.
The mBLOR functional was benchmarked against a variety of other DFT$+U$-type functionals using stretched, homo-nuclear, p-block dimers and was the only corrective functional that 
consistently yielded significantly improved total energies, in this idealized
and hence stringent limit for such functionals,
when compared to the raw DFT (PBE) values. The mBLOR functional was also the only DFT$+U$ functional that opened the bandgap of the stretched, spin-symmetric N$_2$ and F$_2$ molecules, 
as well as H$_2$, when
its explicit derivative discontinuity is `potentialized', 
a technique that may prove useful in other contexts.
Much further study will be needed to understand
how the mBLOR (and its progenitor BLOR) functionals perform
in applied simulation, both for total-energy and 
GKS eigenspectrum based properties.
Its effect on the potential, in particular, appears to 
typically be greater in general than from its BLOR or 
DFT+U counterparts, due to the inclusion of inter-orbital
contributions, and this warrants exploration. 
Yet the present results are most encouraging, and in general
our findings highlight the diagnostic
potential of using exact results that hold
for idealized yet physical test systems, as well
as the promising route of building expedient correctors 
for approximate DFT 
using exact conditions such as the tilted-plane condition.

\section{Acknowledgments}
The research conducted in this publication was funded
by the Irish Research Council under grant number
GOIPG/2020/1454. All calculations were performed on
the Boyle and Kelvin clusters maintained by the Trinity Centre for
High Performance Computing. The Boyle cluster was funded
through grants from the European Research Council \&
Science Foundation Ireland and the Kelvin cluster was funded by the Irish National Infrastructure Federation.

\section{Appendix I: The Asymmetric-MSIE function}
\begin{figure}
\centering
\includegraphics[scale=0.4]{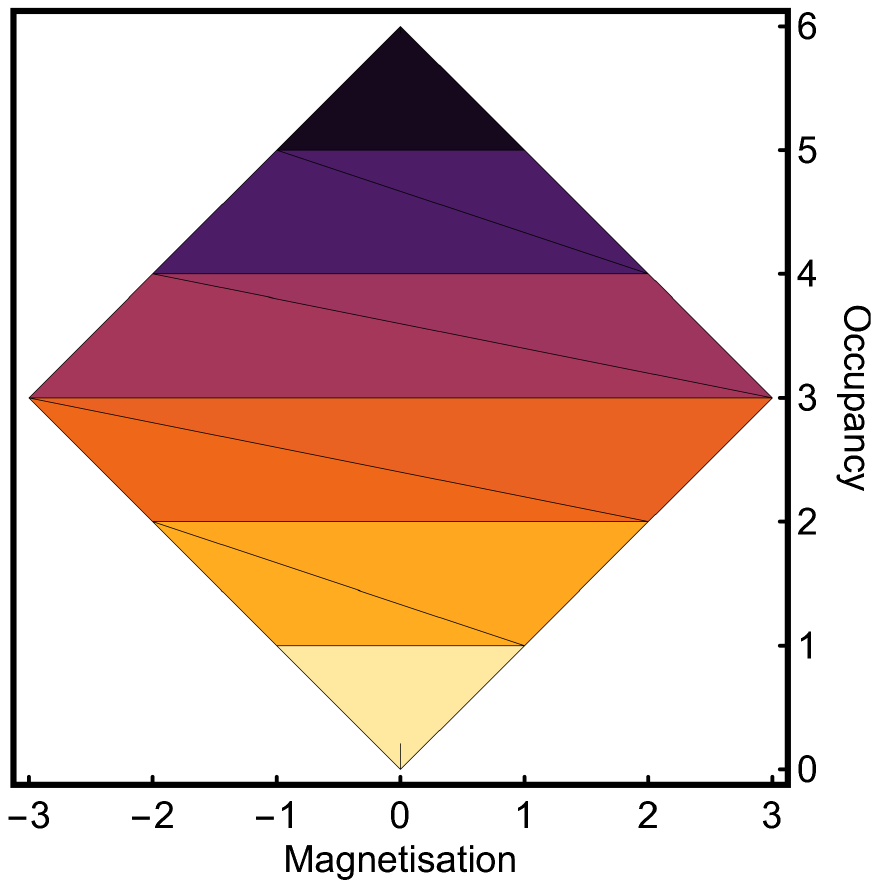}
\caption{The fractured $E_v[N,M]$ energy surface for a p-orbital subspace that satisfies criteria (a)-(c), with $U^{\upharpoonright}>U^{\downharpoonright}$. For ease of visualization, each $N_0 \leq N \leq N_0+1$ segment is highlighted a different color. A single line of fracture occurs within each $N_0 \leq N \leq N_0+1$ segment, with the exception of the $0 \leq N \leq 1$ and $2{\rm Tr}[\hat{P}]-1 \leq N \leq 2{\rm Tr}[\hat{P}]$ segments. The mirror image, through the vertical $M=0$ line, of this fracturing pattern will occur if $U^{\upharpoonright}<U^{\downharpoonright}$, while no diagonal fracturing
occurs for $U^{\upharpoonright}=U^{\downharpoonright}$.
Over-simplistically, but perhaps helpfully 
for visualization of the shown 
$U^{\upharpoonright}>U^{\downharpoonright}$ example, 
we can envisage the greater discretized energy
curvature in the spin-up direction (which
points at 45 degrees up-and-right away from the origin)
necessitating additional fracturing along the vertex
connection lines that are near-perpendicular to that direction.}
\label{fig:p_block_fractured}
\end{figure}
The fracturing pattern for a p-orbital subspace
(as an example) that satisfies criteria (a)-(c), as explicated in the derivation of the mBLOR functional is displayed in Fig.~\ref{fig:p_block_fractured}. The Asymmetric-MSIE function $F_{\rm AMSIE}[N,M]$, is an explicit function of the subspace occupancy and magnetization but it has eight different forms. The following three criteria can be used to determine which form of the AMSIE function should be employed for a given subspace.
\begin{enumerate}
\item 
The relative magnitude of $U^{\upharpoonright}$ and $U^{\downharpoonright}$. A different version of the AMSIE function should be employed if $U^{\upharpoonright}>U^{\downharpoonright}$ or vice-versa.
\item 
If the subspace occupancy is less than half occupied, i.e., $N\leq {\rm Tr}[\hat{P}]$, the `early' version of the AMSIE function should be employed. If the subspace is more than half occupied, i.e., $N>{\rm Tr}[\hat{P}]$, the `late' version of the AMSIE function is required.
\item 
The `upper' or `lower' version of the AMSIE function should be employed depending whether the point $(N,M)$ on the $N$-$M$ plane, where $N$ is the subspace occupancy and $M$ the subspace magnetization, is located above or below the line of fracture within a given $N_0 \leq N \leq N_0+1$ segment. The `upper' version should be employed if it is located above the line of fracture and the `lower' version should be employed if it is located below the line of fracture. No line of fracture occurs in the segment  $0 \leq N \leq 1$, in this case the `upper' version should always be employed, similarly no line of fracture occurs in the segment  $2{\rm Tr}[\hat{P}]-1 \leq N \leq 2{\rm Tr}[\hat{P}]$ and in this case the `lower' version should always be employed.
\end{enumerate}

The eight different forms of the $F_{\rm AMSIE}[N,M]$ function are listed in table~\ref{table:Famsie}. The different forms of the spin up and spin down Asymmetric-MSIE potential operator are listed in table~\ref{table:vamsie}.

\begin{widetext}
\begin{center}
\begin{table}
\centering
\begin{tabular}{|l|} 
 \hline
 $U^{\upharpoonright}>U^{\downharpoonright}$, lower  \\ 
  \hline
 $F_{\rm AMSIE}^{\rm early}\left[N,M\right]=-(N-N_0)(1+N_0+M)$ \\ 
 $F_{\rm AMSIE}^{\rm late}\left[N,M\right]= -(N-N_0)(2{\rm Tr}[\hat{P}]-N_0-1+M)$ \\ 
 \hline 
  $U^{\upharpoonright}>U^{\downharpoonright}$, upper  \\ 
   \hline
 $F_{\rm AMSIE}^{\rm early}\left[N,M\right]=-(N_0+1-N)(N_0-M)$ \\ 
  $F_{\rm AMSIE}^{\rm late}\left[N,M\right]=-(N_0+1-N)(2{\rm Tr}[\hat{P}]-N_0-M)$ \\ 
  \hline  
  $U^{\upharpoonright}<U^{\downharpoonright}$, lower  \\ 
   \hline
 $F_{\rm AMSIE}^{\rm early}\left[N,M\right]=(N-N_0)(1+N_0-M)$ \\ 
 $F_{\rm AMSIE}^{\rm late}\left[N,M\right]=(N-N_0)(2{\rm Tr}[\hat{P}]-N_0-1-M)$ \\ 
 \hline 
  $U^{\upharpoonright}<U^{\downharpoonright}$, upper  \\ 
   \hline
 $F_{\rm AMSIE}^{\rm early}\left[N,M\right]=(N_0+1-N)(N_0+M)$ \\ 
 $F_{\rm AMSIE}^{\rm late}\left[N,M\right]=(N_0+1-N)(2{\rm Tr}[\hat{P}]-N_0+M)$ \\ 
 \hline  
\end{tabular}
\caption{  {The eight different forms of the Asymmetric-MSIE function, $F_{\rm AMSIE}\left[N,M\right]$.}}
\label{table:Famsie}
\end{table}
\end{center}

\begin{center}
\begin{table}%[H]
\centering
\begin{tabular}{|l|l|} 
 \hline
$  \hat{v}^{\upharpoonright}_{\rm AMSIE}$ & $ \hat{v}^{\downharpoonright}_{\rm AMSIE}$
\\
 \hline  
  $U^{\upharpoonright}>U^{\downharpoonright}$, lower  & $U^{\upharpoonright}>U^{\downharpoonright}$, lower  \\ 
   \hline
 $\hat{v}_{\rm AMSIE}^{\rm early}\left[N,M\right]=-\hat{P}-2{N}^{\upharpoonright}\hat{P}$  &  $\hat{v}_{\rm AMSIE}^{\rm early}\left[N,M\right]=(-2N_0-1)\hat{P}+2{N}^{\downharpoonright}\hat{P}$ \\ 
 $\hat{v}_{\rm AMSIE}^{\rm late}\left[N,M\right]=(2N_0+1-2{\rm Tr}[\hat{P}])\hat{P}-2{N}^{\upharpoonright}\hat{P}$  &  $\hat{v}_{\rm AMSIE}^{\rm late}\left[N,M\right]=(1-2{\rm Tr}[\hat{P}])\hat{P}+2{N}^{\downharpoonright}\hat{P}$ \\ 
 \hline 
  $U^{\upharpoonright}>U^{\downharpoonright}$, upper   & $U^{\upharpoonright}>U^{\downharpoonright}$, upper   \\ 
   \hline
 $\hat{v}_{\rm AMSIE}^{\rm early}\left[N,M\right]=(1+2N_0)\hat{P}-2{N}^{\upharpoonright}\hat{P}$  & $\hat{v}_{\rm AMSIE}^{\rm early}\left[N,M\right]=-\hat{P}+2{N}^{\downharpoonright}\hat{P}$ \\ 
  $\hat{v}_{\rm AMSIE}^{\rm late}\left[N,M\right]=   (1+2{\rm Tr}[\hat{P}])\hat{P}-2{N}^{\upharpoonright}\hat{P}$  &   $\hat{v}_{\rm AMSIE}^{\rm late}\left[N,M\right]=  (2{\rm Tr}[\hat{P}]-2N_0-1)\hat{P}+2{N}^{\downharpoonright}\hat{P} $ \\ 
 \hline 
 $U^{\upharpoonright}<U^{\downharpoonright}$, lower & $U^{\upharpoonright}<U^{\downharpoonright}$, lower \\ 
  \hline
 $\hat{v}_{\rm AMSIE}^{\rm early}\left[N,M\right]=(1+2N_0)\hat{P}-2{N}^{\upharpoonright}\hat{P}$  &  $\hat{v}_{\rm AMSIE}^{\rm early}\left[N,M\right]=\hat{P}+2{N}^{\downharpoonright}\hat{P}$ \\ 
 $\hat{v}_{\rm AMSIE}^{\rm late}\left[N,M\right]=(2{\rm Tr}[\hat{P}]-1)\hat{P}-2{N}^{\upharpoonright}\hat{P}$  &  $\hat{v}_{\rm AMSIE}^{\rm late}\left[N,M\right]= (2{\rm Tr}[\hat{P}]-2N_0-1)\hat{P}+2{N}^{\downharpoonright}\hat{P}$ 
 \\ 
 \hline 
  $U^{\upharpoonright}<U^{\downharpoonright}$, upper  & $U^{\upharpoonright}<U^{\downharpoonright}$, upper  \\ 
   \hline
 $\hat{v}_{\rm AMSIE}^{\rm early}\left[N,M\right]=\hat{P}-2{N}^{\upharpoonright}\hat{P}$  & $\hat{v}_{\rm AMSIE}^{\rm early}\left[N,M\right]=(-2N_0-1)\hat{P}+2{N}^{\downharpoonright}\hat{P}$ \\ 
  $\hat{v}_{\rm AMSIE}^{\rm late}\left[N,M\right]=(1+2N_0-2{\rm Tr}[\hat{P}])\hat{P}-2{N}^{\upharpoonright}\hat{P}$  &   $\hat{v}_{\rm AMSIE}^{\rm late}\left[N,M\right]= (-2{\rm Tr}[\hat{P}]-1)\hat{P}+2{N}^{\downharpoonright}\hat{P}$ \\ 
 \hline  
\end{tabular}
\caption{The eight different forms of the spin resolved Asymmetric-MSIE potential operator.}
\label{table:vamsie}
\end{table}
\end{center}

\end{widetext}
\hspace{20cm}

\hspace{20cm}

\section{Appendix II: Computational details}

All calculations were performed using the ONETEP (Order-N Electronic Total Energy Package) DFT code \cite{prenticeONETEPLinearscalingDensity2020,skylarisIntroducingONETEPLinearscaling2005,skylarisNonorthogonalGeneralizedWannier2002,oreganLinearscalingDFTFull2012}. The ONETEP code constructs the 
Kohn-Sham 
density matrix $\rho({\bf r},{\bf r}')$ from a set of Non-orthogonal Generalized Wannier Functions (NGWFs) $\{ \phi_{\alpha} \}$,
\begin{equation}
\rho({\bf r},{\bf r}')=\sum_{\alpha, \beta}\phi_{\alpha}({\bf r})K^{\alpha \beta}\phi_{\beta}({\bf r}'),
\end{equation}
where $K^{\alpha \beta}$ is the density kernel. The total energy of the system is minimized by optimizing both $K^{\alpha \beta}$ and $\{ \phi_{\alpha} \}$.

All calculations were completed using the PBE \cite{perdewGeneralizedGradientApproximation1996} exchange-correlation functional at a high cutoff energy of no lower than $1$,$500$ eV. The dissociated molecular test systems were located in a large simulation cell, no smaller than $70 \times 60 \times 60$ $a_0^3$, with a Martyna-Tuckerman periodic boundary correction cutoff of $7.0$ $a_0$ \cite{martynaReciprocalSpaceBased1999}.

For a system with $N_{\sigma}$ spin $\sigma$  {GKS} particles, the occupancy of the lowest $N_{\sigma}$  {GKS} particles was set equal to one, and otherwise set equal to zero. The convergence threshold of the root-mean-square gradient of the density kernel and the NGWFs was set at $1\times 10^{-6}$ Ha $e^{-1}$ and $1\times 10^{-7}$ Ha $a_0^{3/2}$, respectively, and the electronic energy tolerance was set at $1\times 10^{-6}$ eV/atom.

Seven NGWFs were assigned per atom using the split-valence approach
for p orbitals, in which case  $15\%$ of the norm set to be beyond the matching radius $r_m$. 
The NGWF cutoff was set to $14$ $a_0$. A bespoke set of norm-conserving pseudopotentials with very small cut off radii were made using the OPIUM code \cite{OPIUM}. 
It is important to emphasise that the corrective parameters as well as the subspace occupancies depend strongly on the choice of subspace projection operator $\hat{P}$. For our purposes we use the atomic orbitals generated using the Pseudoatomic Solver in ONETEP with our bespoke norm-conserving pseudopotentials,
specifically the pseudoatomic p-orbitals generated 
in a neutral non-spin-polarized atomic reference state
within the PBE approximation.
In the limit of large interatomic separation lengths, the atomic orbitals become an ideal choice for the subspace projection operator, particularly when using PBE atomic reference energies
as part of the energy extensivity diagnostic. Thus, by testing the Hubbard functionals on molecular species at large interatomic separation lengths, any resulting errors in the total energies can be attributed to failures in the underlying DFT$+U$-type functional as opposed to failures in the subspace projection scheme. 

In the case of the stretched singlet N$_2$, F$_2$ and spin-polarized O$_2$ molecules the raw PBE calculation converged to a spurious spin symmetry broken solution. This results in spin resolved PBE densities that differ qualitatively from the true ground state spin resolved densities of the molecule. This necessitates the evaluation of the Hubbard corrective parameters self-consistently  {(in the specific sense that
we proceed to describe)} as the spin resolved densities will change appreciably upon application of a Hubbard type corrective functional. 

In the case of the stretched F$_2$ molecule a series of linear response perturbative calculations were used to evaluate the Hubbard corrective parameters ($U_{\rm out}$ and $J_{\rm out}$) with Dudarev's DFT$+U$ functional applied to stabilise the spin symmetry unbroken ground state. The corrective parameters $U_{\rm out}$ and $J_{\rm out}$ were evaluated at a range of values of $U_{\rm in}$ between $-18$ eV and $-22$ eV, where $U_{\rm in}$ is the input value of the Hubbard $U$ parameter used in Dudarev's DFT$+U$ functional to stabilise the spin symmetry unbroken ground state. The self-consistent Hubbard $U$ and Hund's $J$ corrective parameters were found by linearly extrapolating the values of $U_{\rm out}$ and $J_{\rm out}$ back to $U_{\rm in}=$ 0 eV. The same procedure was also implemented for the stretched N$_2$ and O$_2$ molecules. However, for these systems the DFT+$U$+$J$ functional without the minority spin term was used to stabilise the  symmetry unbroken ground state, with the effective Hubbard $U_{\rm in}$ parameter set equal in magnitude to the Hund's $J_{\rm in}$ parameter. $U_{\rm out}$ and $J_{\rm out}$ were evaluated at a range of values of $U_{\rm in}$ between $-8$ eV and $-12$ eV in the case of the N$_2$ molecule and between $-21$ eV and $-24$ eV in the case of the O$_2$ molecule. 

Despite the  {authors'} best efforts, no suitable range in values of $U_{\rm in}$ were found for the O$_2$ molecule in its spin polarized state that sufficiently stabilized the symmetry unbroken solution to allow linear extrapolation of $U_{\rm out}(U_{\rm in})$ back to $U_{\rm in}=$ 0 eV. As a result, the calculations on the O$_2$ molecule were performed in its non-spin polarized state as a suitable range in values of $U_{\rm in}$ could be found for this system. It was worth noting that the spin polarized and non-spin polarized states of the O$_2$ molecule should be degenerate in energy as they represent the $m_s=1$ and $m_s=0$ values of the molecule in its triplet ground state. In many regards the non-spin polarized state of the stretched O$_2$ molecule poses a significantly more challenging test case for the mBLOR functional as unlike the spin polarized state, the non-spin polarized state will be located at the point of maximum localized-SCE and being a neutral homonuclear molecule at large separation lengths, their will be negligible localized-MSIE present. The values of the Hubbard corrective parameters for each stretched molecular species is reported in table~\ref{table:hubbard_parameters}.

\begin{widetext}
\begin{center}
\begin{table}
\centering
\begin{tabular}{|c|c|c|c|c|c|c|} 
 \hline  
   & $U^{\upharpoonright}$ & $U^{\downharpoonright}$& $U_{\rm simple}$ & $J_{\rm simple}$& Magnetization &  Separation Length\\ 
   \hline
   H$_2$ & 6.783 & 6.783 &  8.689 & 1.905 & 0.0 & 9.0\\
      \hline
   He$_2^+$ & -37.961 & 13.729 &  -13.837 & -1.721 & 1.0& 5.0\\
   \hline
    Li$_2$ & 7.507 & 7.507 &  9.251 & 1.743 & 0.0& 15.0\\
   \hline
    Be$_2^+$ & -11.881 & 3.988 &  -4.238 & -0.291 & 1.0& 10.0\\
   \hline
    H$_5^{+}$ & 9.827 & 4.536 &  9.154 & 1.972 & 2.0& 8.0\\
   \hline
    N$_2$ & 7.450 & 7.450 &  8.189 & 0.740 & 0.0& 7.0\\
   \hline
    O$_2$ & 8.156 & 8.156 &  9.037 & 0.881 & 0.0& 6.0\\
   \hline
      F$_2$ & 10.471 & 10.471 &  11.429 & 0.958 & 0.0& 6.0\\
   \hline
      Ne$_2^+$ & -43.872 & 12.855 &  -17.384 & -1.875 & 1.0 & 5.0\\
   \hline
\end{tabular}
\caption{The Hubbard and Hund corrective parameters (eV) for the stretched dimers,  evaluated using the minimum tracking linear response methodology  {(within the 
simple $2 \times 2$ scheme that effectively constrains
$N$ while $M$ varies, and vice versa)}. 
The total magnetization $M$ 
(unitless electron count)
and inter-nuclear separation length in Bohr radii ($a_0$) is also reported for each molecular species.
For completeness and comparison, we also report the values of the 
first-principles corrective parameters for the stretched s-block species that were evaluated for testing the BLOR corrective functional~\cite{burgessMathrmDFTTexttypeFunctional2023}.}
\label{table:hubbard_parameters}
\end{table}
\end{center}
\end{widetext}

\bibliography{main}% Produces the bibliography via BibTeX.

\end{document}